\documentclass{article}
\usepackage[utf8]{inputenc}
\usepackage{graphicx}

\usepackage{amsmath}
\usepackage{amssymb}
\usepackage{appendix}
\usepackage{siunitx}
\usepackage{float}
\usepackage{hyperref}
    \hypersetup{
        colorlinks=true,
        linkcolor=blue,
        filecolor=blue,      
        urlcolor=blue,
        citecolor=blue
    }
\usepackage[left=4cm, right=4cm, top=2.5cm, bottom=2.5cm]{geometry}

\usepackage{subfigure}

\title{Pedestrian Flow Analysis in High-Density Crowds: Continuity Equation with Voronoi-Based Fields}

\author{Juliane Adrian, Ann Katrin Boomers, Sarah Paetzke, Armin Seyfried}
\date{June 19, 2025}

\graphicspath{{1DFD/}}

\begin{document}

\maketitle

\textbf{Abstract.} 
Since the beginning of the century, capturing trajectories of pedestrian streams precisely from video recordings has been possible. To enable measurements at high density, the heads of the pedestrians are marked and tracked, thus providing a complete representation of the phase space. However, classical definitions of flow, density, and velocity of pedestrian streams are based on different segments in phase space. In addition, traditional methods fail with high densities of people, as heads move even when a crowd is blocked and standing still. In this article, Voronoi decomposition is used to construct density and velocity fields from pedestrian trajectories to solve this problem. Combined with the continuity equation, a flow equation on the basis of trajectories is derived satisfying the conservation of particle numbers exactly. 
The proposed method allows definitions of all quantities in the same segment of phase space even on scales smaller than the dimensions of a pedestrian. It is shown that these new definitions of flow, density, velocity are consistent with classical measurements and make it possible to determine standstill in pedestrian flows even when individual body parts are moving. These properties allow to scrutinize inconsistencies in the state of the art of pedestrian fundamental diagrams.

\section{Introduction}
\label{sec:intro}

The density, flow, velocity, or speed are helpful in quantitatively describing collective phenomena and transport properties of a crowd. They are utilised as variables for modelling crowd dynamics but also to characterize states with respect, e.g., to risks and comfort. In addition, these quantities are empirically interrelated. The relationship between these quantities gained by empirical measurements is referred to as a fundamental diagram. Different representations, such as speed-density, flow-density or flow-speed are used. These relationships facilitate the categorisation of system states, such as the congested or free-flow regime. In addition, the fundamental diagram specifies states that are also relevant for safety in a crowd. In particular, it allows determining the density or the density regime in which congestion builds up or at which movement comes to a standstill and crowding, clogging, or pushing can occur.

Irrespective of the empirical relationships, various concepts or models are used in theory that link speed $v$, density $\rho$, and flow $J$ through functions or conservation laws. In pedestrian dynamics, the flow equation $J=\rho\; v\; w$, where $w$ gives the width of the facility, is used in every textbook \cite{PRED78ENG, Fruin1971, Buchmueller2007, Nelson2002} and in applications \cite{PED01, PED16}. In general, these quantities are related to discrete representations of the system, usually the trajectories describing the movement of the pedestrians. Even though the flow equation is widely used in practice, it raises fundamental questions. For example, it links quantities such as the flow, an average value over time at a fixed cross-section in space, with the density, usually an average value over space at a fixed point in time. These quantities, therefore, relate to different areas in phase space. In mathematics or physics, many macroscopic models are based on the continuity equation as well as on model functions for the fundamental diagram, see 
\cite{Henderson1974, Hughes2002} or the review \cite{bellomo_modeling_2011}. Here, density $\boldsymbol{\rho}(\vec{x},t)$ and velocity $\vec{\mathbf{v}}(\vec{x},t)$ are space- and time-dependent continuous fields. The continuity equation is a conservation law describing changes of density and velocity in time and space, ensuring that no pedestrian in the system is born or turns to dust.

To summarise, the following could be stated. There are different relationships between the central variables to describe transport and safety risks in crowds: (a) empirical measurements of speed, density, and flow as well as the fundamental diagram; (b) the flow equation and (c) the continuity equation describing transport in moving crowds while conserving the number of pedestrians. However, there is as yet no standard in the scientific community for the definition of measured and model variables. This article will show that these inaccuracies and uncertainties can lead to the fact that important states in crowds are described and interpreted incorrectly. In the following, we introduce the state of the art concerning data capturing and empirical measurements of density, velocity, and flow, consider the completeness and inconsistencies in the determination of the fundamental diagram, and discuss how these inconsistencies are related to the continuity or the flow equation.


Prior to the turn of the millennium, quantitative data collection predominantly depended on manual methods (e.g., \cite{Hankin1958, Hoel1968}) or labor-intensive analysis of photographic and video materials (e.g., \cite{navin_pedestrian_1969, tanaboriboon_pedestrian_1986}). The emergence of high-resolution video recording technology for the consumer market significantly facilitated the detailed and efficient documentation of experiments and field studies through video recordings. These recordings enable not only a qualitative analysis but also an automated or half-automated analysis of pedestrians' movement, e.g., \cite{als_Hoogendoorn05a, Seyfried2007d}. To solve the problem of occlusion in experiments with a high density, overhead recordings are made. The head is usually marked in colour with a kind of cap and is used for detecting and tracking pedestrians \cite{hoogendoorn2003extracting}. The position of the head is then projected onto the ground, providing two-dimensional trajectories. In order to allocate pedestrians in space correctly, the perspective distortion at the edge of the video recordings has to be considered. To implement this, the subjects wear caps marked with colour or aruco-codes for the height of the person \cite{Boltes2010} or stereo recordings \cite{Boltes2015} are used. The accuracy depends on the position of the head in the camera image as well as details of the methodology. It is 0 cm directly below the camera and typically reaches a magnitude of 10 cm at the edge of the image. Details on the factors influencing this error can be found in \cite{Boltes2016a}. These precise head trajectories formed the basis for microscopic analysis, the development of new measurement methods and representations of density, speed, and flow, as well as the influence of the measurement method on the quantities and their statistical properties \cite{Helbing2007, Steffen2010, Zhang2011, CaoS2017, Cao2018}. 

In this article it will be shown that this new precision in trajectories highlights weaknesses and ambiguities in the classic definitions of key variables used to describe transport properties in crowds. To address these problems, the article introduces a framework that allows the definition of the variables density, flow, and velocity (speed) as well as the determination of the fundamental diagram in consistency with the continuity equation. 
 
Subsection \ref{Variables} introduces a microscopic and macroscopic representation of the variables based on head trajectories. In Subsection \ref{ContinuityEquation}, the continuity equation is used to relate direct measurements of the flow to a flow calculated by the flow equation, which is in accordance with the continuity equation. To connect the continuity equation, a field equation, with the discrete representation of moving pedestrians by trajectories, the Voronoi decomposition is used. In subsections \ref{Unifloweq} and \ref{Bifloweq}, the framework is then used to introduce a flow equation in accordance with the continuity equation and tested by analyzing experiments of unidirectional and bidirectional streams.

\section{State of the art for measuring transport properties}

As outlined at the beginning the fundamental diagram is of central importance to describe transport in crowds. In this section open questions and contradictions in the state of the art are pointed out. In the second part of the section it is discussed how these problems relate to the definition of variables.

\subsection{Fundamental diagram}

Many empirical studies have shown that the fundamental diagrams depend on human factors such as body height, gender, culture, and many more. Also, the influence of anisotropy of the moving direction (bidirectional or multidirectional streams) and the type of spatial boundaries (corridors, intersections, stairs, etc.) was studied. For an overview on this subject we refer to \cite{Boltes2018} or the collection of empirical works in \cite{Haghani2020, Feng2021}. Many studies, including the works of the authors, analyse these influences by comparing fundamental diagrams that are not based on the same measurement method. First, problems in this context were discussed in \cite{Seyfried2010a} for single file movement. This study used microscopic trajectories from one experiment to measure mean values of density, speed, and flow. The flow equation was used to compare different representations of the fundamental diagram. However, the comparison obtained by the flow equation results in inconsistent shapes of the fundamental diagram. The flow equation relates mean values over space at a fixed time, like the density, with mean values over time at a fixed position, like the flow. This leads to discrepancies in the relationship between the empirical measurements at high densities, where the heterogeneity of the system, e.g., due to stop-and-go waves, is pronounced. In the following, we focus on the incompleteness of the data and inconsistencies resulting from the definition of speed, density, and flow. Second, further studies on the influence of the measurement method of speed, density, and flow based on trajectories revealed a significant influence on the shape and the fluctuation of the data also for systems with larger degrees of freedom to move \cite{Zhang2011, CaoS2017, Cao2018}. 

Without any claim to completeness, we next focus on data from experiments under laboratory conditions on single-file, uni- and bi-directional flow in straight corridors. The collection includes studies that cover the density range as complete as possible. All experiments differ in the spatial boundaries of the setup, the test persons, and the initial conditions at the start of the experiment. Only experiments with no additional obstacles in the corridor and not using further accessories such as mobile phones or luggage are considered.

\begin{figure}[htp]
    \centering
    \subfigure[]{
    \includegraphics[width=0.475\textwidth]{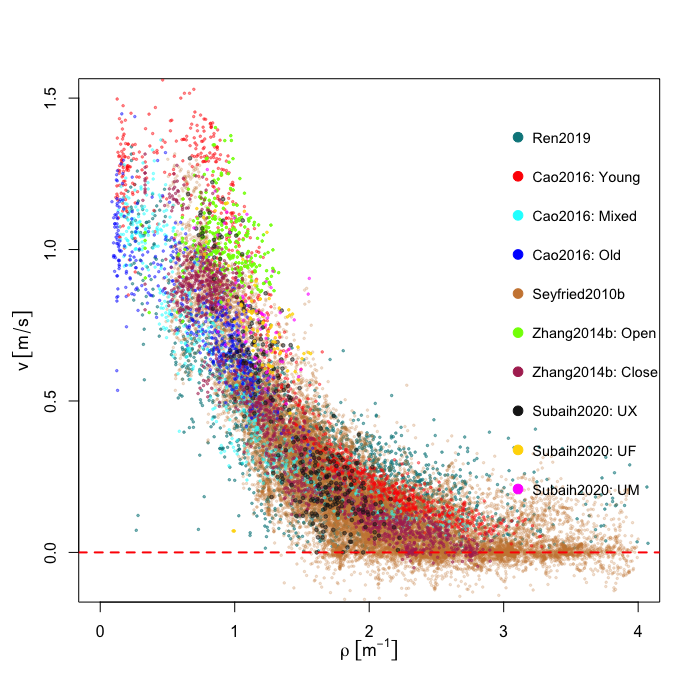}
    }
    \subfigure[]{
    \includegraphics[width=0.475\textwidth]{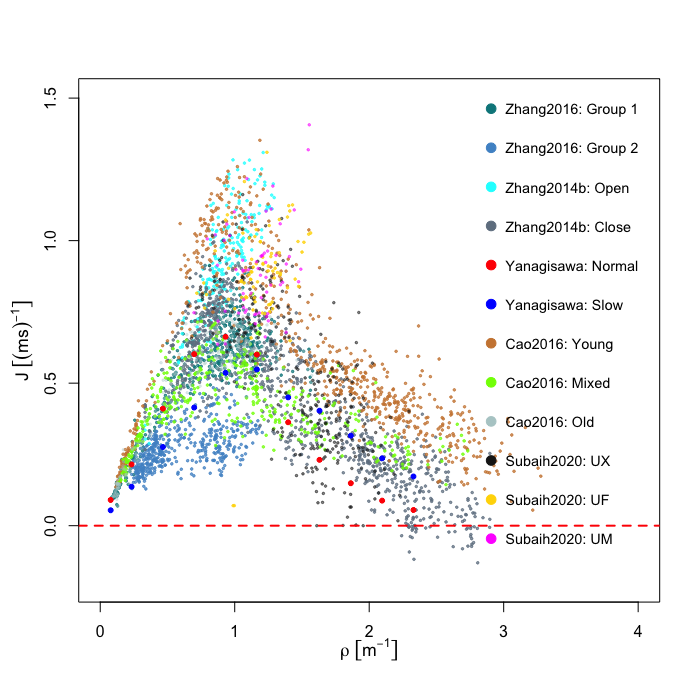}
    }
    \caption{Fundamental diagrams for single-file experiments, a) speed-density, b) flow-density.} 
    \label{fig:singlefile_literature_data}
\end{figure}

\autoref{fig:singlefile_literature_data} shows data from various single file experiments \cite{ZhangJ2016, Ren2019a, Subaih2020, ZiemerDiss, Jelic2012, CaoS2016, Subaih2019, CaoS2018, CaoS2019, Zhang2014b, Zhang2014c}. The scope of this collection is not to discuss how the human factors, the experimental layout or the instruction influence the fundamental diagram. The compilation shows that, regardless of these optional influencing factors, the flow of people can come to a complete standstill at high densities, and the speed can even reach negative values. This is also in line with expectations. At high densities, there is no more room for movement, and taking steps is no longer possible. If pedestrians no longer move forward and stand in one place, this does not mean that the head no longer moves. The negative velocities result from the measurement of the head movement and the balance shifts caused by the change of the stance leg. This phenomenon is also expected in experiments on unidirectional or bidirectional flows in wide corridors. The lateral movement offers more room for manoeuvre, but at a certain density, a standstill must nevertheless occur. 

\begin{figure}[h!]
    \centering
    \subfigure[]{\includegraphics[width=0.475 \textwidth]{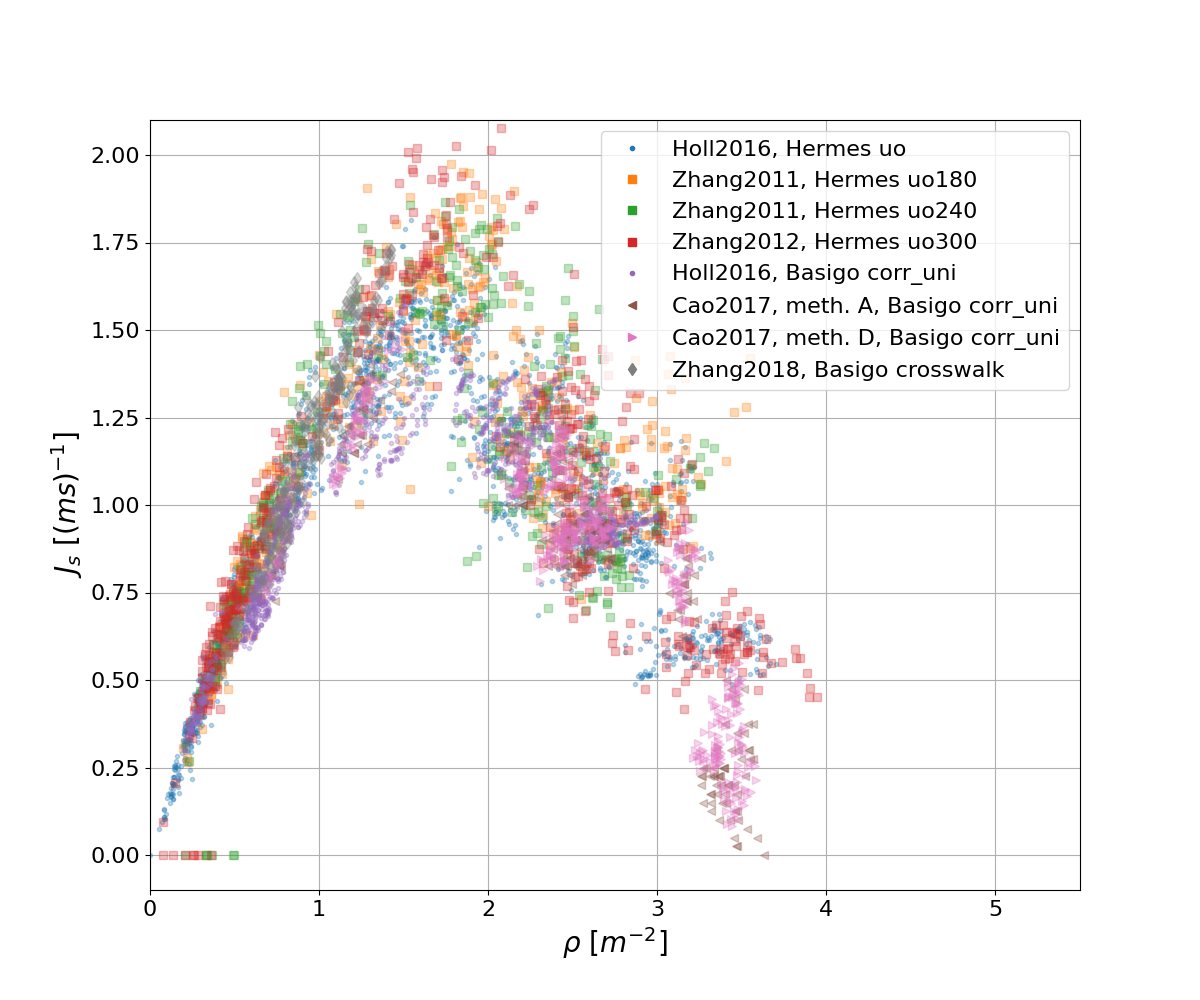}}
    \label{fig:corridor_literatureData_a}
    \subfigure[]{\includegraphics[width=0.475 \textwidth]{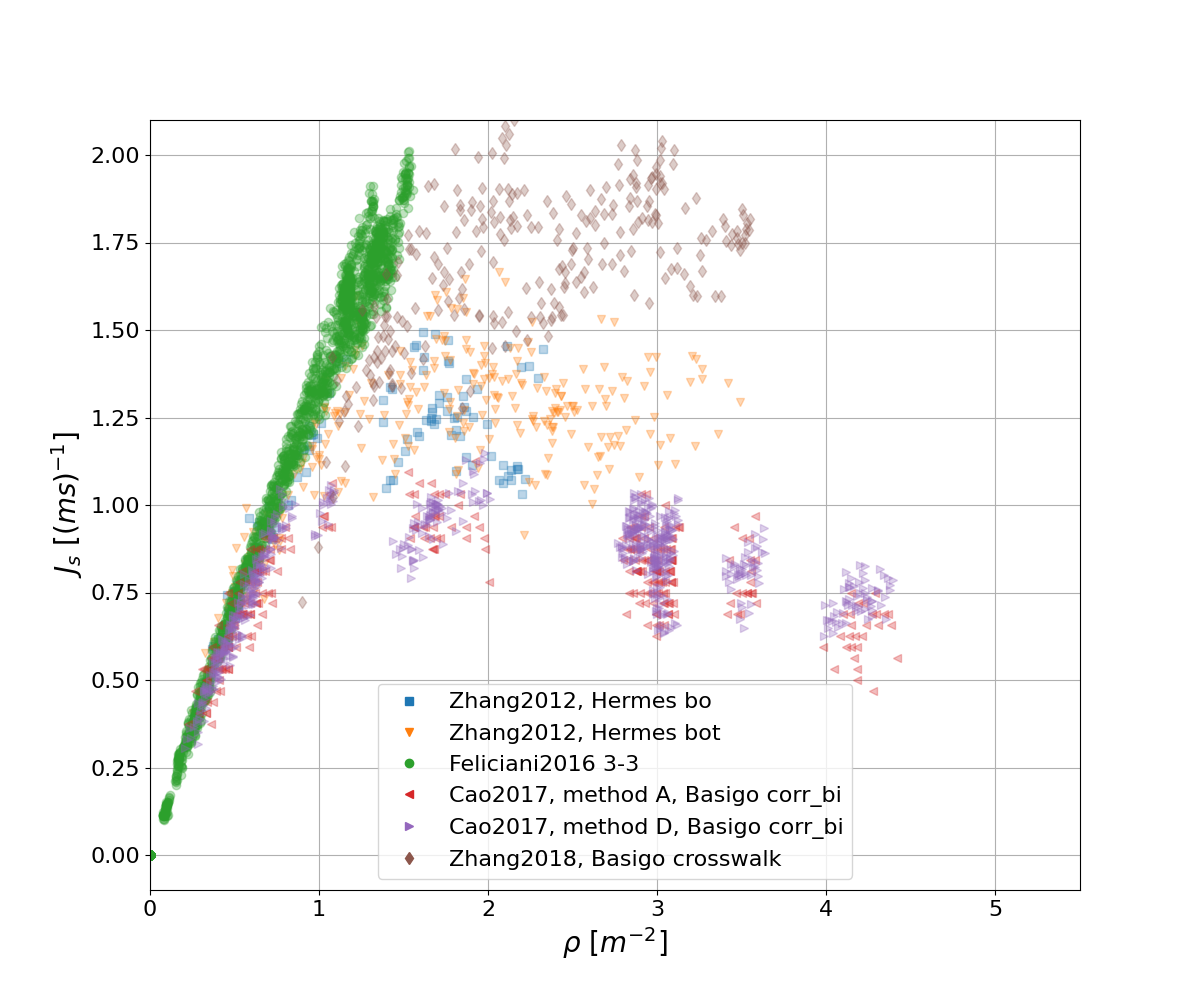}}
    \label{fig:corridor_literatureData_b}
    \subfigure[]{\includegraphics[width=0.475 \textwidth]{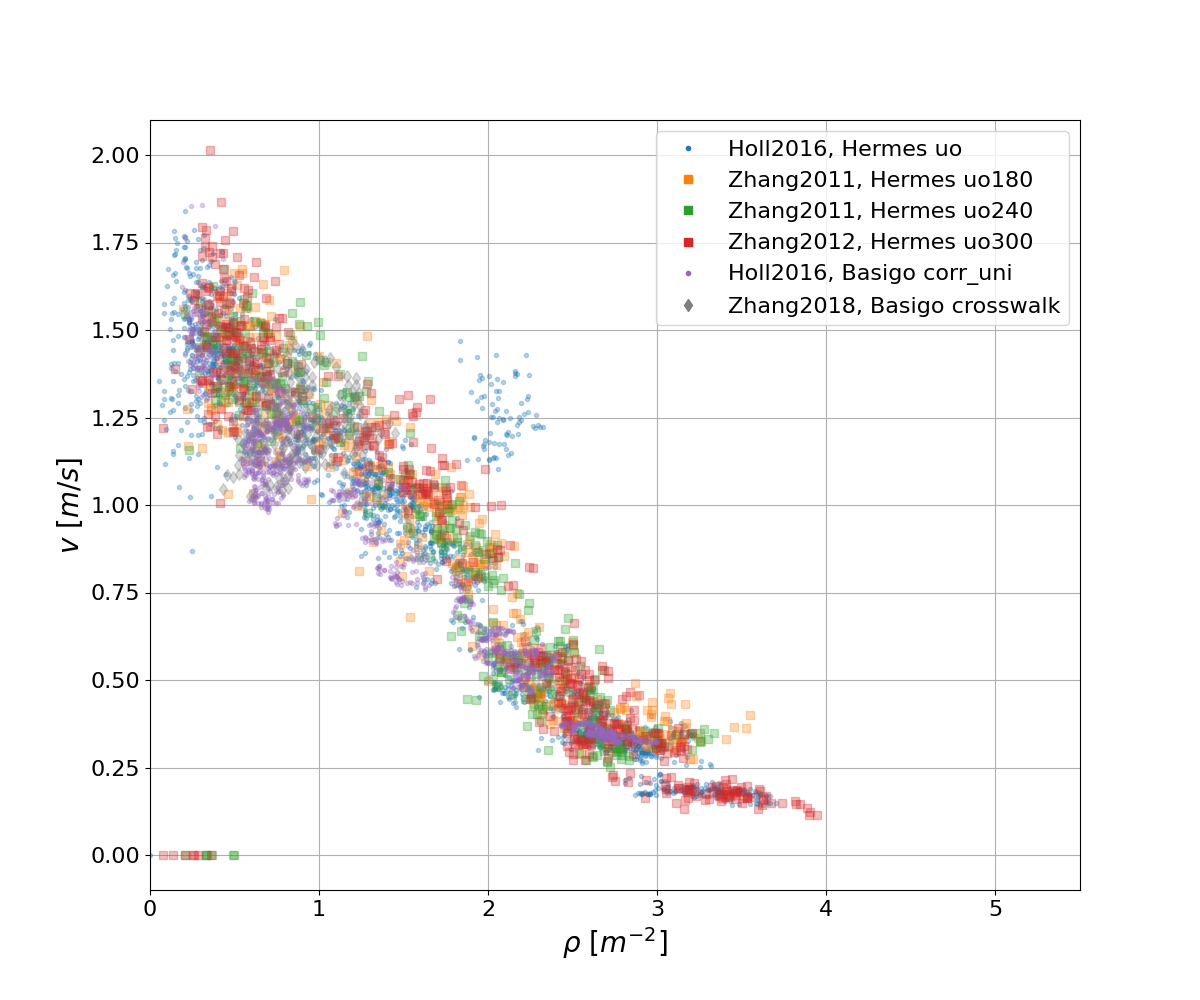}}
    \label{fig:corridor_literatureData_c}
    \subfigure[]{\includegraphics[width=0.475 \textwidth]{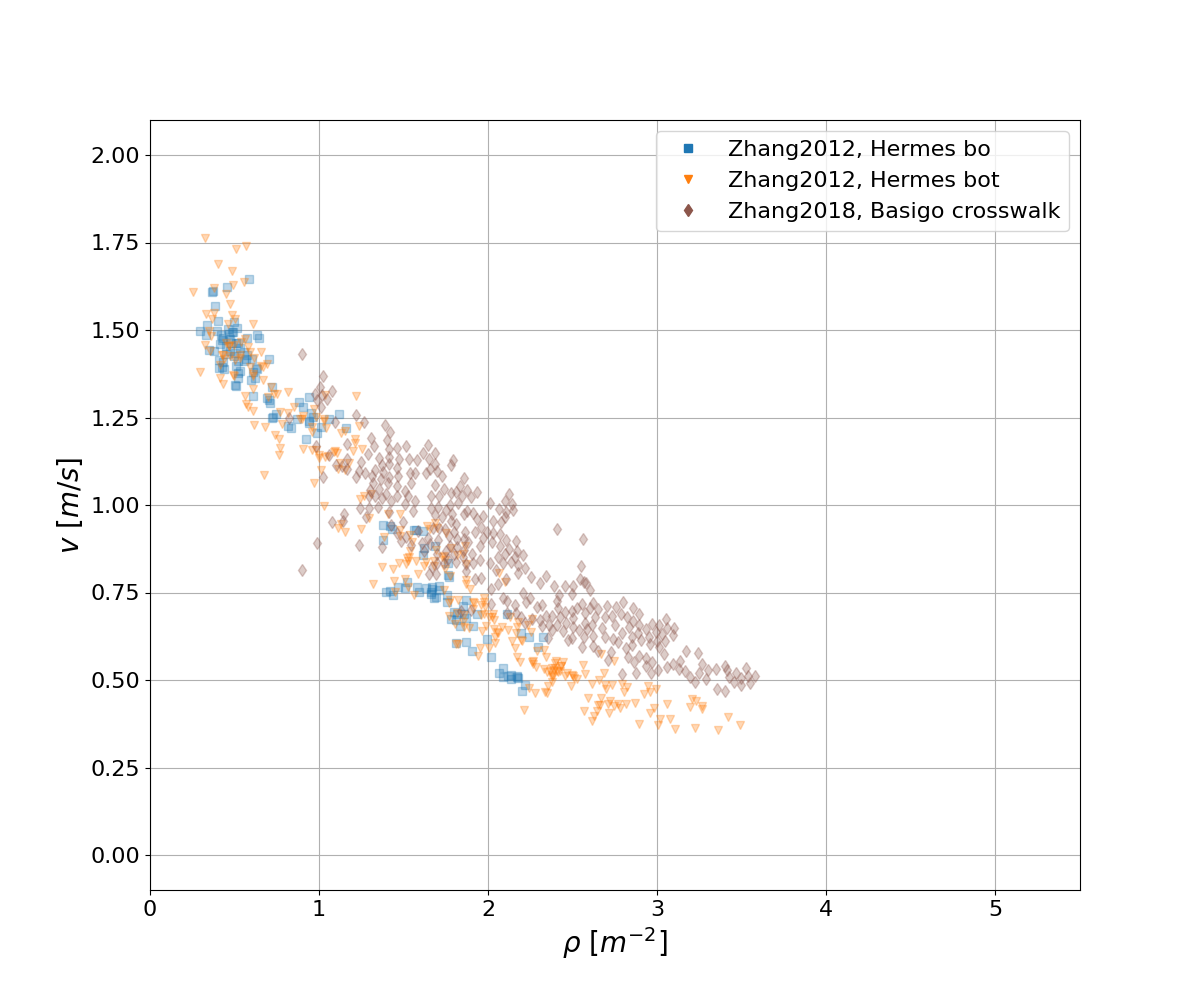}}
    \label{fig:corridor_literatureData_d}
    \caption{Fundamental diagrams for corridor experiments. a)+c) Unidirectional streams, b)+d) bidirectional streams with two stationary lanes and bidirectional streams with multiple intersecting bands.}
    \label{fig:corridor_literatureData}
\end{figure}

Three different types of stream directions in a corridor are considered in the following: First, uni-directional streams \cite{Zhang2011, Zhang2012, Holl2016, CaoS2017, Zhang2018a}, second, bi-directional streams with two clearly separated stationary lanes \cite{Zhang2012, Feliciani2016, CaoS2017} and third, bi-directional streams with multiple intersecting bands changing in space and time \cite{Zhang2018a, Zhang2012, Feliciani2016}. The respective data is shown in \autoref{fig:corridor_literatureData}. Again, data collection includes experiments covering the density range as completely as possible. 
The shape of the presented fundamental diagrams differs greatly between unidirectional and bidirectional movement and between the studies itself, as has been widely recognised. Also, they differ greatly in the measurement methods of how the variables for the fundamental diagram are obtained. On the other hand they differ in the method how density and flow are obtained. Both quantities can be either directly measured \cite{CaoS2017} or calculated \cite{Zhang2011, Feliciani2016, CaoS2017, Holl2016, Cao2018}. Among the later studies, flow is calculated by density and speed using the flow equation, while the density calculation is based on \cite{Steffen2010, Zhang2011}. Some of the most recent studies show that the need to precisely define which portions of the respective quantities add to the flow or which sign has to be applied has been recognised. Most important, even under controlled conditions in the laboratory, the expected standstill at high densities does not appear. It is also noticeable that, unlike in experiments on single file movement, all speed and flow values are positive.

\subsection{Variables}
\label{Variables}

\begin{figure}[htp]
            \centering
            \includegraphics[width=10cm]{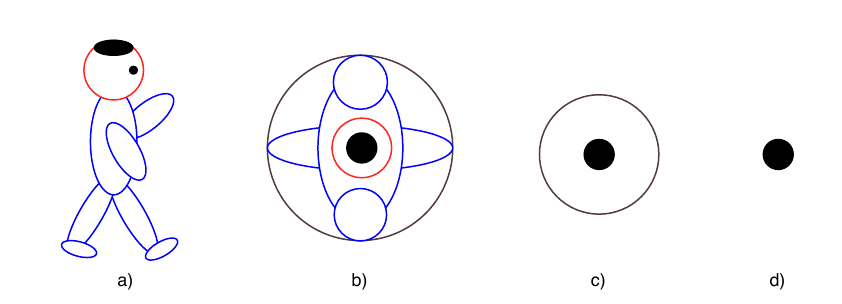}
            \caption{Simplification of the movement of a pedestrian for the state of the art description of transport properties of crowds. a) shows the pedestrian from a side view. In b), this is shown from the top view. While some approaches consider a volume exclusion, e.g., by a circle c) most analysis is based on a point indicating the position of the head d).} \label{fig:body}
        \end{figure}

The movement of a pedestrian is performed by a three-dimensional body that continuously changes its shape by walking in steps. The powerful locomotive system is based on a pendulum movement of the legs and uses arms and body, for example, to maintain balance or rotate the upper body to pass through narrow spaces. To enable the analysis of many pedestrians moving in a crowd, the definitions of density, speed, and flow are based on a strong simplification. In general, the position of a body part, usually the head, is projected onto a two-dimensional space and the head, with a diameter of \SI{0.2}{m}, is represented by a point (\autoref{fig:body}).

\begin{figure}[htp]
    \centering
    
    \subfigure[]{
        \begin{minipage}{0.375\textwidth}
            \includegraphics[trim=0 0 0 0, clip, width=1\textwidth]{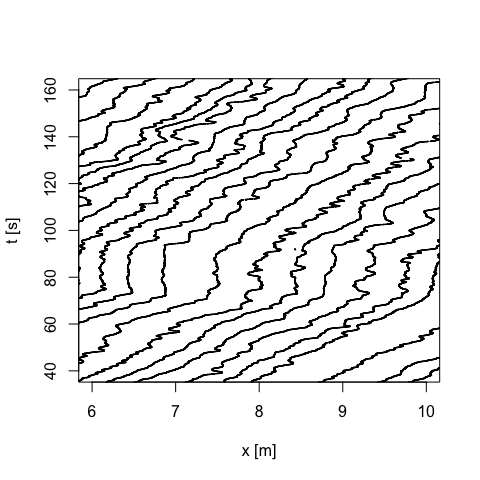}
            \includegraphics[trim=0 0 0 0, clip, width=1\textwidth]{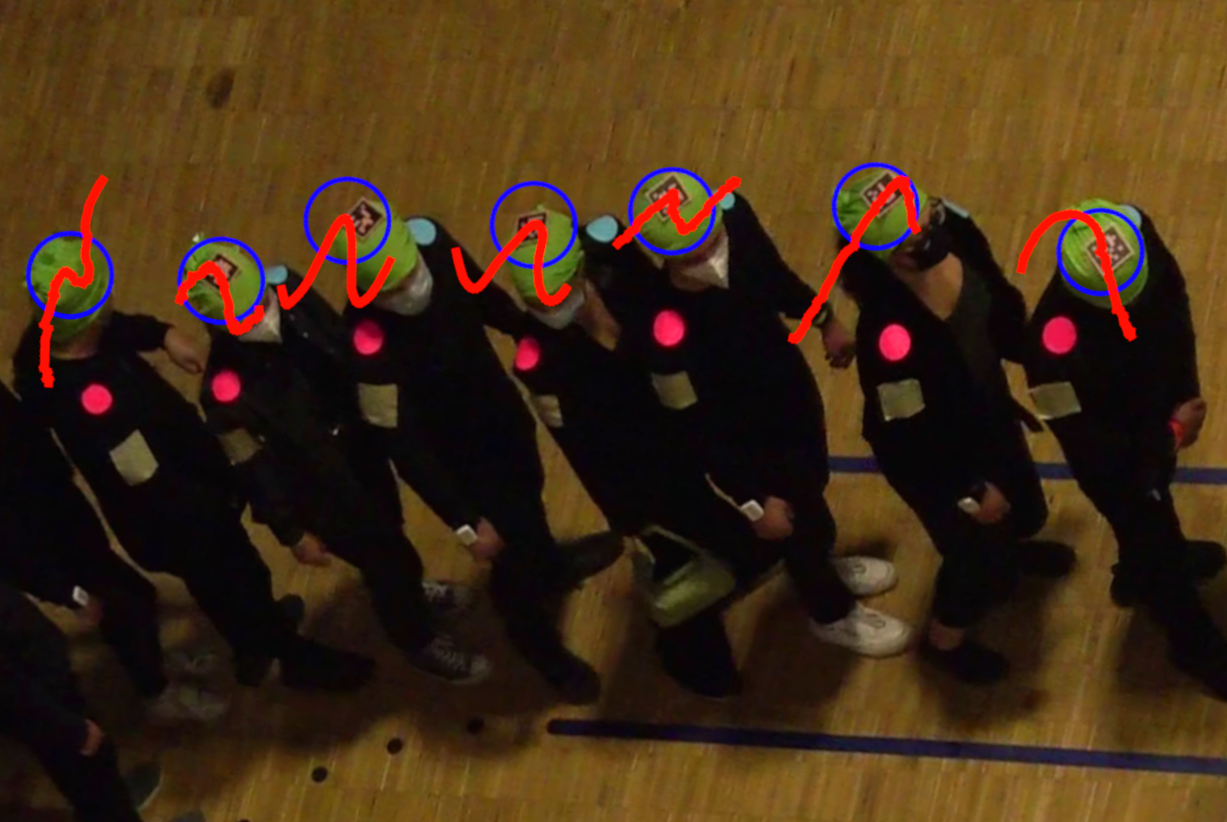}   
        \end{minipage}
        \label{fig:trajcap_a}
    }
    \subfigure[]{
        \begin{minipage}{0.465\textwidth}
            \includegraphics[trim=32 25 30 0, clip, width=1\textwidth]{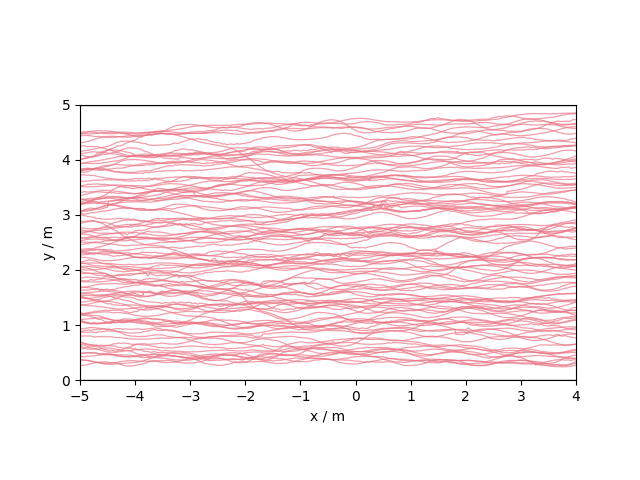}
            \includegraphics[trim=0 0 0 0, clip, width=0.97\textwidth]{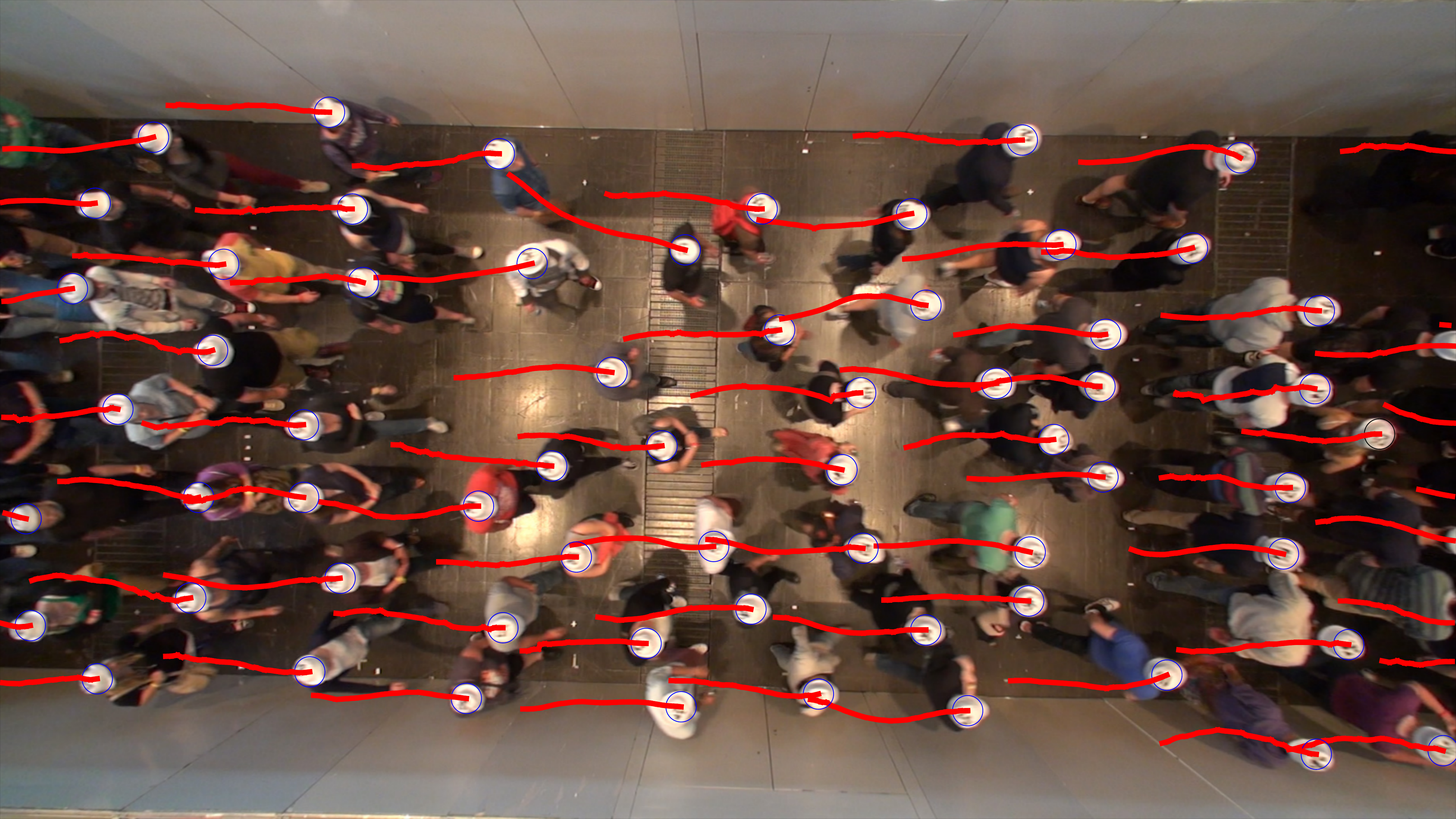}
        \end{minipage}
        \label{fig:trajcap_b}
    }

    \caption{a) Example experiment single-file. Top: individual trajectories with $x$ being the direction of movement. Bottom: Snapshot of the experiment. b) Example experiment unidirectional flow in a corridor. Top: individual trajectories. Bottom: Snapshot of the experiment. In both snapshots, blue circles show the current head positions. Red lines indicate the head trajectory in a time interval $[\SI{-1}{s}, \SI{1}{s}]$ in (a) and $[\SI{-1}{s}, \SI{0}{s}]$ in (b) from the current frame.}
    \label{fig:trajcap}
\end{figure}

For the definition of variables describing transport properties of crowds, the movement of the human body of person $i$ in time $t$, performed in a three dimensional world, is abstracted by the movement of a point in two dimensions with position $\vec r_i(t)$ (see \autoref{fig:trajcap}).

\subsubsection*{Microscopic representation}

The movement of individual person $i$ could be described by the position vector $\vec{x}_i(t)$ and the velocity $\vec{v}_i(t)$:

\begin{equation}
    \vec{x}_i(t)=\left( \begin{array}{ccc} x_i(t) \\ y_i(t) \end{array} \right)
        \quad
    \vec{v}_i(t)=\dot{\vec{x}}_i(t)=\left( \begin{array}{ccc} \dot{x}_i(t) \\ \dot{y}_i(t) \end{array} \right).
    \label{eq:variables_position_micro}
\end{equation} 

The position vector $\vec{x}_i(t)$ in time is the trajectory. Even for single-file movement, the trajectory gained by tracking the head of a pedestrian is two dimensional. The bipedal movement of the human body causes the upper body and head to sway orthogonally to the main direction of movement, in Figure \ref{fig:trajcap_a} the direction $x$. At high densities, pedestrians may come to a stop, standing in one place but shifting their balance from one leg to the other, causing their heads to move despite standing still. As a representation of such a situation by trajectories in the $x$-$t$-plane shows, the head may move in the opposite direction to the main direction of movement $x$, which must lead to negative velocities. For pedestrian streams in wider corridors, the main direction of movement could be two-dimensional, i.e. in the $x$-$y$-plane (cf. Figure\,\ref{fig:trajcap_b}). 

Two ways are used to define the speed. The magnitude $v_i$ of the velocity vector or the component of the velocity in the main moving direction $\vec{n}_{main}$:

\begin{equation}
 v_i (t) = \mid \vec{v}_i(t) \mid  \quad \text{or} \quad  v_{i,main}(t) = \vec{v}_i(t) ~\vec{n}_{main}.
\end{equation}

While position and velocity can be assigned to a single person, quantities like distance or flow need to be related to at least a second person $j$.

\begin{figure}[htp]
    \centering
    \subfigure[]{
        \includegraphics[width=.29\textwidth]{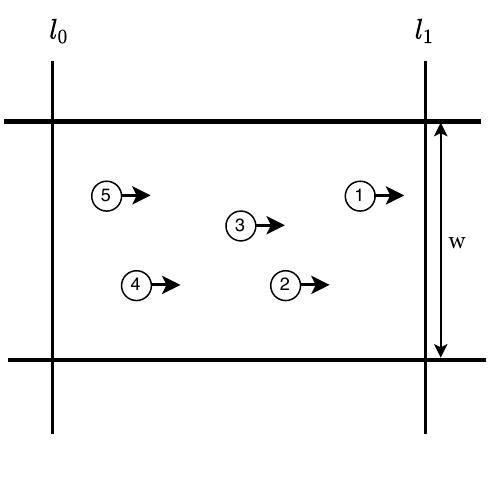}
    \label{fig:variables_trajectories+measurementline_a}}
    \subfigure[]{
        \includegraphics[width=.29\textwidth]{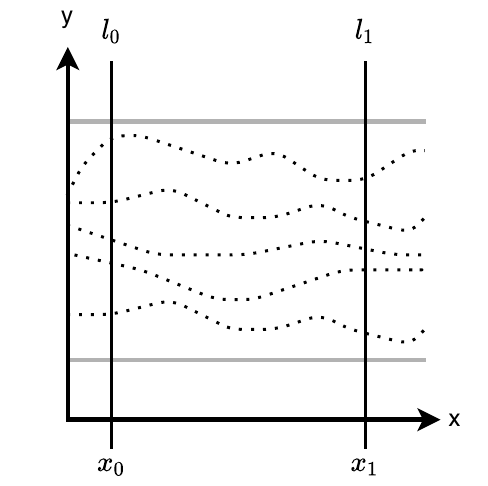}
        \label{fig:variables_trajectories+measurementline_b}}
    \subfigure[]{
        \includegraphics[width=.33\textwidth]{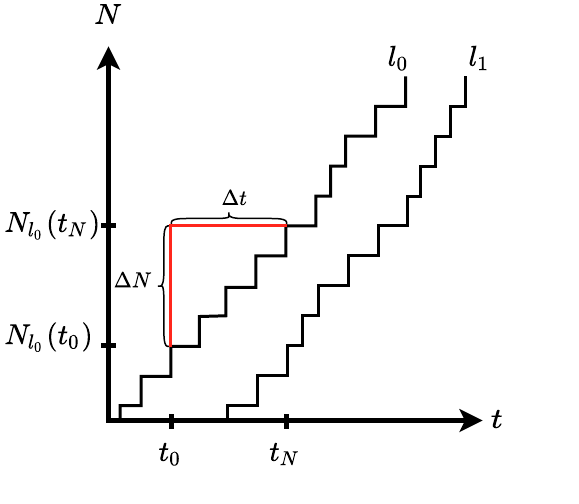}
    \label{fig:variables_trajectories+measurementline_c}}
    \caption{a) Sketch of a corridor with width $w$ in which five individuals walk from left to right between measuring lines $l_0$ and $l_1$, b) trajectories of individuals moving on the x-y plane in x-direction between $l_0$ at $x=x_0$ and $l_1$ at $x=x_1$, c) $N-t$ diagram showing the cumulative count $N$ of people having crossed $l_0$ and $l_1$ at time $t$.}
    \label{fig:variables_trajectories+measurementline}
\end{figure}

To measure the flow, a cross-section, e.g., a line $l_0$ at $x_0$, has to be introduced, see \autoref{fig:variables_trajectories+measurementline_a}. The time at which the pedestrians cross the line $l_0$ is then $t_{i}(x_0)=t_{i,l_0}$. These times are sorted from the first to the last crossing to determine the time interval between two consecutive people crossing $l_0$ which is $\tau_{i,l_0}=t_{i-1}(x_0)-t_{i}(x_0)$. Following the above, an individual microscopic flow can be assigned to a person $i$ by $J_{i,l_0}=(\tau_{i,l_0})^{-1}$. 

The size of the area that is assigned to an individual person is $A_i$. The value $A_i$ can either be estimated by the inverse of the macroscopic density or by the size of the Voronoi cell associated with the pedestrian $i$ \cite{Steffen2010}. 

For a free software developed for analysing pedestrian trajectories, we refer to \href{https://pedpy.readthedocs.io/en/v1.1.2/}{PedPy}, \cite{schrodter_2024_11778532}.

\subsubsection*{Macroscopic representation}

In Physics and Engineering, the classical way for a macroscopic description of pedestrian systems is based on mean values over time and space using trajectories, in other words, the Lagrange specification of the system. In the following, we use superscript $c$ to index these variables. In a mathematical context, variables could also be based on an Euler specification of the systems by fields. While later the Voronoi-decomposition is used to define the fields, variables based on fields are indexed by the superscript $\text{v}$. 

For mean values in time, the position is fixed at measurement line $l_0$, and the movement is regarded within the time interval $\Delta t$. The start and end points of the time interval used for averaging are assigned to the crossing times of the first pedestrian $t_{0,l_0}$ and the last pedestrian $t_{N,l_0}$, respectively. $N_{l_0,\Delta t}$ then indicates the number of people who have crossed the measurement line $l_0$ in the time interval $]t_{0,l_0},t_{N,l_0}]$, see Figure \autoref{fig:variables_trajectories+measurementline_c}. The flow is then given by

\begin{equation}
    J_{l_0,\Delta t}^c = \dfrac{N_{l_0,\Delta t} }{\Delta t + \epsilon(t)} ,
\label{eq:variables_flow_macro}
\end{equation}

where $\epsilon$ considers that for a fixed magnitude of the time interval $\Delta t$, the time of crossing of the last person is not necessarily equal to the of time crossing of the first person plus the time interval selected for the measurement: $t_{N,l_0}\neq t_{0,l_0} + \Delta t$.
For simplicity, $\Delta t + \epsilon$ is denoted by $\Delta t$ in the following. Figure  \autoref{fig:variables_trajectories+measurementline_c} shows that introducing $\epsilon$ is necessary to consider the discreteness of crossings. A further possibility to handle this problem is implemented in the software PedPy \cite{schrodter_2024_11778532}, see the documentation in the \href{https://pedpy.readthedocs.io/en/v1.1.2/user_guide.html#flow-at-bottleneck}{userguide}. Another detail is the handling of crossings of the line against the main direction of movement, which can be caused by head movements when pedestrians stand, for example. In this case, the last crossing in the direction of movement is usually counted.

To assign a temporal mean value of the velocity to the macroscopic flow at $l_0$, individual velocities $\vec{v}_i(t_{i,l_0})$ at the time of the transition $t_{i,l_0}$ over the measuring line are calculated and averaged. The mean velocity within the time interval $\Delta t$ at line $l_0$ can be obtained by

\begin{equation}
    \vec{v}_{l_0,\Delta t}=\dfrac{1}{N_{l_0,\Delta t}} \sum_{i=1}^{N_{l_0,\Delta t}}\vec{v}_i(t_{i,l_0}).
    \label{eq:variables_<v>_x0}
\end{equation}

For the mean values of the speed over time, the magnitude or the component of the main moving direction, here the direction orthogonal to the measurement line $\vec{n}_{l_0}$ is used:

\begin{equation}
    v_{l_0,\Delta t}^c=\dfrac{1}{N_{l_0,\Delta t}} \sum_{i=1}^{N_{l_0,\Delta t}} v_{i,l_0} \quad \text{or} \quad v_{l_0,\Delta t,\perp}^c=\dfrac{1}{N_{l_0,\Delta t }} \sum_{i=1}^{N_{l_0,\Delta t }}v_{i,\perp}.
    \label{eq:variables_<v>_x0_perp}
\end{equation}
  
For the calculation of mean values of the velocity in space, the measurements are based on an area $A$ while the time is fixed at $t_0$, e.g. between both measurement lines $l_0$ and $l_1$ in Figure\,\ref{fig:variables_trajectories+measurementline_b}. The mean velocity $\vec{v}_{t_0, A}$ within the measurement area $A = \Delta x  \cdot w$ with $\Delta x = x_1-x_0$, $w$ being the width of the corridor in the y-direction, can be obtained by 

\begin{equation}
        \vec{v}_{A,t_0}=\dfrac{1}{N_{A,t_0}} \sum_{i=1}^{N_{A,t_0}}\vec{v}_i(t_0),
        \label{eq:variables_<v>_t0} 
\end{equation}

where $N_{A,t_0}$ is the number of pedestrians inside $A$ at time $t_0$. Again, it should be differentiated between mean values of the magnitude $v_{A,t_0}^c$ or the velocity component in the main moving direction $v_{A,t_0,\perp}^c$. Classically, the density $\rho$ is defined as 

\begin{equation}
    \rho_{A}^c(t)=\dfrac{N_{A}(t)}{A},
    \label{eq:density_classical}
\end{equation}    

where $N_{A}(t)$ is the number of pedestrians inside $A$ at time $t$. Unfortunately, the size of the objects (here, the size of the pedestrians) is of the same order of magnitude as the size of $A$, leading to high fluctuations of the classical density. To overcome that problem, a Voronoi decomposition of the space could be used, see \cite{Steffen2010}. The positions $\vec{x}_i(t)$ of pedestrians $i=1,...,N$ are used to construct a scalar density field ${\boldsymbol{\rho}}(\vec{x},t)$ by decomposing the space into Voronoi cells $A_i(t)=A_i(\vec{x}_i(t))$ based on the positions of the pedestrians $\vec{x}_i(t)$. The resulting field is given by 

\begin{align}
    {\boldsymbol{\rho}}(\vec{x},t) = \sum_i \boldsymbol{\rho}_i(\vec{x},t) ~ \text{with} ~ \boldsymbol{\rho}_i(\vec{x},t) &= \left\{\begin{array}{ll} 1/A_i(t) : & \vec{x}\in A_i(t) \\
    0: & otherwise\end{array}\right\}.
\end{align}

The Voronoi decomposition is only one possibility to construct a density field. Other options include the Gaussian distribution \cite{johansson2008}. We prefer the Voronoi decomposition as it is parameter-free and allows easy consideration of boundaries. The density in a measurement area $A$ at a certain time $t$ is then

\begin{equation}
    \rho_{A}^{\text{v}}(t) = \frac{1}{A} \int_A {\boldsymbol{\rho}}(\vec{x},t) ~ d\vec{x}.
    \label{eq:voronoi_density}
\end{equation}

Using the distribution functions to define a density field smoothes the density measurements in space and time. It allows to define an individual density $\rho_{i}(t)= \rho_{A_i(t)}^{\text{v}} = 1/A_i(t)$ or even the definition of measurement areas smaller than the size of a single pedestrian.

\section{Continuity equation with Voronoi based fields}

The microscopic and macroscopic definitions of the quantities in the previous section form the basis for measurements of the fundamental diagram and the application of the flow equation. Various aspects point to fundamental problems in the use of the variables with these definitions. An inspection of which type of mean value is linked already reveals inconsistencies. For example, $J(\rho)$ relates $J_{l_0,\Delta t}$ to $\rho_{A,t_0}$ and thus combines the mean value over time with the mean value over space. Thus, the quantities relate to different areas in phase space. 
The same inconsistency appears in the flow equation $J = \rho\,v\,w$. In \cite{Seyfried2010a}, it was shown which inconsistencies can occur when converting fundamental diagrams using the flow equation. But even if mean values over time and space are used separately and are not mixed, the discussion in section \ref{sec:intro} shows that until now, the fundamental diagrams are not in conformance with plausible expectations. In relation to the particle number conservation, which is manifested by the continuity equation, the following problems of the flow equation arise. First, classical measurements of the mean values are not uniformly definable with respect to the measurement area in space and time. Second, the formulation is not sufficiently precise for the calculation of $v$. The velocity of individual pedestrians $\vec{v}_i$ is a vector in a two dimensional system, including information about the direction of the movement and a definition of the scalar mean value $v=f(\vec{v}_1(t), ...,\vec{v}_N(t))$ is generally not provided. 

There are two historical reasons for this shortcoming. On the one hand, the research area is strongly oriented to the research area of vehicular traffic, which is treated as a quasi-one-dimensional system due to the traffic lanes. On the other hand, historical data are associated with very rough estimates of speeds, flows, and densities, which make a precise definition of $v$ seemingly unnecessary. However, with the capture of individual two-dimensional trajectories in a time discreteness of about a tenth of a second, accuracy is achieved, which may become relevant, especially for high densities and inhomogeneous systems. Moreover, the used trajectories represent the head position while the head of pedestrians is sometimes associated with large fluctuations and velocities in the opposite direction as the walking direction results in additional problems. 

\subsection{Continuity equation}
\label{ContinuityEquation}

In the following, the field formulation of the continuity equation is the starting point to enable a spatio-temporally consistent measurement of the scalars $j,\rho$ and $v$. To link this formulation with the measurement of discrete trajectories of single pedestrians $\vec{x}_i(t)$, the Voronoi decomposition $A_i(\vec{x}_i(t))$ is used to construct the fields. Firstly, the flow equation is formulated for a typical measurement scenario:

\begin{equation}
\frac{\mathrm{d}N_A}{\mathrm{d}t} + \oint_l \vec{\boldsymbol{j}}(\vec{x},t) \cdot \mathrm{d}\vec l = 0, 
\end{equation}

where $N_A$ is the number of all pedestrians in the area $A$, and $l$ is a line enclosing the area. $\vec j$ is the flow field of $N_A$. The flow field is connected with the velocity field $\vec{u}$ by  
\begin{equation}
    \vec{\boldsymbol{j}}(\vec{x},t) = \boldsymbol{\rho}(\vec{x},t) \boldsymbol{\vec{v}}(\vec{x},t).
    \label{eq:j_rho_u}
\end{equation}    
Here, it is important to note that this is an integral of the flow field along a line. When the flow field is directed parallel to the line, the flow through the line is zero. When it flows perpendicular to the line, it is maximum. Only the component of the flow vector which is perpendicular to the line $j^{\perp}=\vec{\boldsymbol{j}} \cdot \vec n$ is contributing to the change of number of pedestrians in $A$, with $\vec n$ being the normal vector to the line. 

For the example represented in Figure \ref{fig:variables_trajectories+measurementline_b} the measurement area $A = w \cdot (x_{1} - x_{0})$ is enclosed by line $l$ consisting of two lines in x-direction along the corridor walls between $l_0$ and $l_1$ and two lines in y-direction along $l_0$ and $l_1$ itself with a length of $w$ each. Since there is no movement through the walls of the corridor, the flow only needs to be measured along the measurement lines $l_0$ and $l_1$. Then, the continuity equation is

\begin{equation}
\frac{\mathrm{d}N_A}{\mathrm{d}t} = \int_{l_0} \vec{\boldsymbol{j}} \cdot \mathrm{d}\vec{l} + \int_{l_1} \vec{\boldsymbol{j}} \cdot \mathrm{d}\vec{l} 
= \int_{l_0} \boldsymbol{\rho} \vec{\boldsymbol{v}} \cdot \vec n  \cdot \mathrm{d}y + \int_{l_1} \boldsymbol{\rho} \vec{\boldsymbol{v}} \cdot \vec n \cdot \mathrm{d}y,
\label{eq:conteq_euler}
\end{equation}

with $\vec{n}$ pointing towards the inside of the measurement area. 
For the flow as mean value within time interval $\Delta t = [t_0,t_N]$ (cf. \autoref{eq:variables_flow_macro}), the continuity equation is then

\begin{equation}
\frac{N_{t_N,A}-N_{t_0,A}}{t_N-t_0} = J_{l_0,\Delta t}+J_{l_1,\Delta t}.
\label{eq:conteq_lagrange}
\end{equation}

Using equation \autoref{eq:j_rho_u}, the flow equation for a certain point in time at line $l_0$ at $x_0$ is 

\begin{equation}
J_{l_0}(t) =
\int_{y_0}^{y_1} \vec{\boldsymbol{j}}(x_0,y,t) ~ \vec n_{l_0} ~ \mathrm{d} y  = \int_{y_0}^{y_1} \boldsymbol{\rho}(x_0,y,t) ~ \vec{\boldsymbol{v}}(x_0,y,t) ~ \vec{n}_{l_0}  ~ \mathrm{d}y.
\label{eq:flowequation_fields_t}
\end{equation}

The macroscopic flow in the time interval $\Delta t$ is given by 

\begin{equation}
J_{l_0,\Delta t} = \frac{1}{\Delta t}
\int_{t_0}^{t_N} \int_{y_0}^{y_1} \vec{\boldsymbol{j}}(x_0,y,t) ~ \vec n_{l_0} ~ \mathrm{d} y ~ \mathrm{d}t = \frac{1}{\Delta t} \int_{t_0}^{t_N} \int_{y_0}^{y_1} \boldsymbol{\rho}(x_0,y,t) ~ \vec{\boldsymbol{v}}(x_0,y,t) ~ \vec{n}_{l_0}  ~ \mathrm{d}y ~ \mathrm{d}t. 
\label{eq:flowequation_fields_deltat}
\end{equation}

Comparing equation \ref{eq:flowequation_fields_deltat} with the flow equation $J=\rho ~ v ~ w$ reveals the following ambiguities: \\

First, the flow equation does not specify how the value of the scalar $v$ is calculated from a velocity, which is a vector. In particular, for the oscillating trajectories of the heads of pedestrians, the calculation of the magnitude of the velocity leads to an overestimation of the fraction of the velocity that contributes to the flow or the throughput. Movements of the heads opposite to the main direction of the motion of the stream, e.g. when pedestrians are standing still, have to be considered by a negative contribution to the velocity and flow. For the calculation of a real throughput in the case of pedestrian streams with different main movement directions (bi- or multidirectional streams), it is then mandatory to distinguish negative contribution due to oscillations of the head opposite to the main direction with positive contributions in the main direction from the opposing stream. 

Second, the density $\rho$ is usually measured as a mean value over a two dimensional measurement area. Only in cases where the density is homogeneous in space it is a reasonable approximation of the density at a measurement line. In particular, at high densities where congestion leads to density waves moving through the measurement area, the difference between a line density and an area density could be significant. 

Third, the multiplication with the length of the measurement line $w$ implies that the product of density and velocity along the measurement line is constant, which is related to the specific flow concept. However, density and velocity or their product could also be inhomogeneous along the measurement line, what can be accompanied by a violation of the concept of specific flow.

Fourth, one has to consider the Cauchy-Schwartz inequality
\begin{equation}
    \int_{l_0} \boldsymbol{\rho} ~ \vec{\boldsymbol{v}} ~ \mathrm{d}y \ne \int_{l_0} \boldsymbol{\rho} ~ \mathrm{d}y \int_{l_0} \vec{\boldsymbol{v}} ~ \mathrm{d}y
\label{eq:cauchy}\\
\end{equation}
meaning that the multiplication of the mean values of the density and speed is not necessarily equal to the mean value of density times speed. 

\subsection{Flow equation by Voronoi fields for unidirectional streams}
\label{Unifloweq}

The following methodology is proposed to handle the problems outlined above and connect the trajectories of pedestrians with the continuity equation. First, the trajectories $\vec{x}_i(t)$ of single pedestrians are used to calculate the Voronoi decomposition and to introduce a density field $\boldsymbol{\rho}$ and a velocity field $\vec{\boldsymbol{v}}$. 

   \begin{align}
    {\boldsymbol{\rho}}(\vec{x},t) = \sum_i \boldsymbol{\rho}_i(\vec{x},t) ~ \text{with} ~ \boldsymbol{\rho}_i(\vec{x},t) &= \left\{\begin{array}{ll} 1/A_i(t) : & \vec{x}\in A_i(t) \\
    0: & otherwise\end{array}\right\}
    \end{align}

   \begin{align}
    \vec{\boldsymbol{v}}(\vec{x},t) = \sum_i \vec{\boldsymbol{v}}_i(\vec{x},t) ~ \text{with} ~ \vec{\boldsymbol{v}}_i(\vec{x},t) &= \left\{\begin{array}{ll} \vec{v}_i(t) : & \vec{x} \in A_i(t) \\
    0: & otherwise\end{array}\right\}
    \end{align}

These fields are used with the flow equation \autoref{eq:flowequation_fields_t} to define the flow field $\vec{\boldsymbol{j}}(\vec{x},t)$ following the continuity equation.  

\begin{align}
J_{l_0}^{\text{v}}(t) &=
\int_{y_0}^{y_1} \vec{\boldsymbol{j}}(x_0,y,t) ~ \vec n_{l_0} ~ \mathrm{d} y \\ &=\int_{y_0}^{y_1} \boldsymbol{\rho}(x_0,y,t) ~ \vec{\boldsymbol{v}}(x_0,y,t) ~ \vec n_{l_0} ~ \mathrm{d} y  \\ &= \int_{y_0}^{y_1} \left( \sum_i \boldsymbol{\rho}_i(x_0,y,t)\right) ~ \left( \sum_j \vec{\boldsymbol{v}}_j(x_0,y,t) ~ \vec{n}_{l_0}\right)  ~ \mathrm{d}y 
\label{eq:abc}
\end{align}

\begin{figure}[htp]
    \centering
    \includegraphics[width=.8\textwidth]{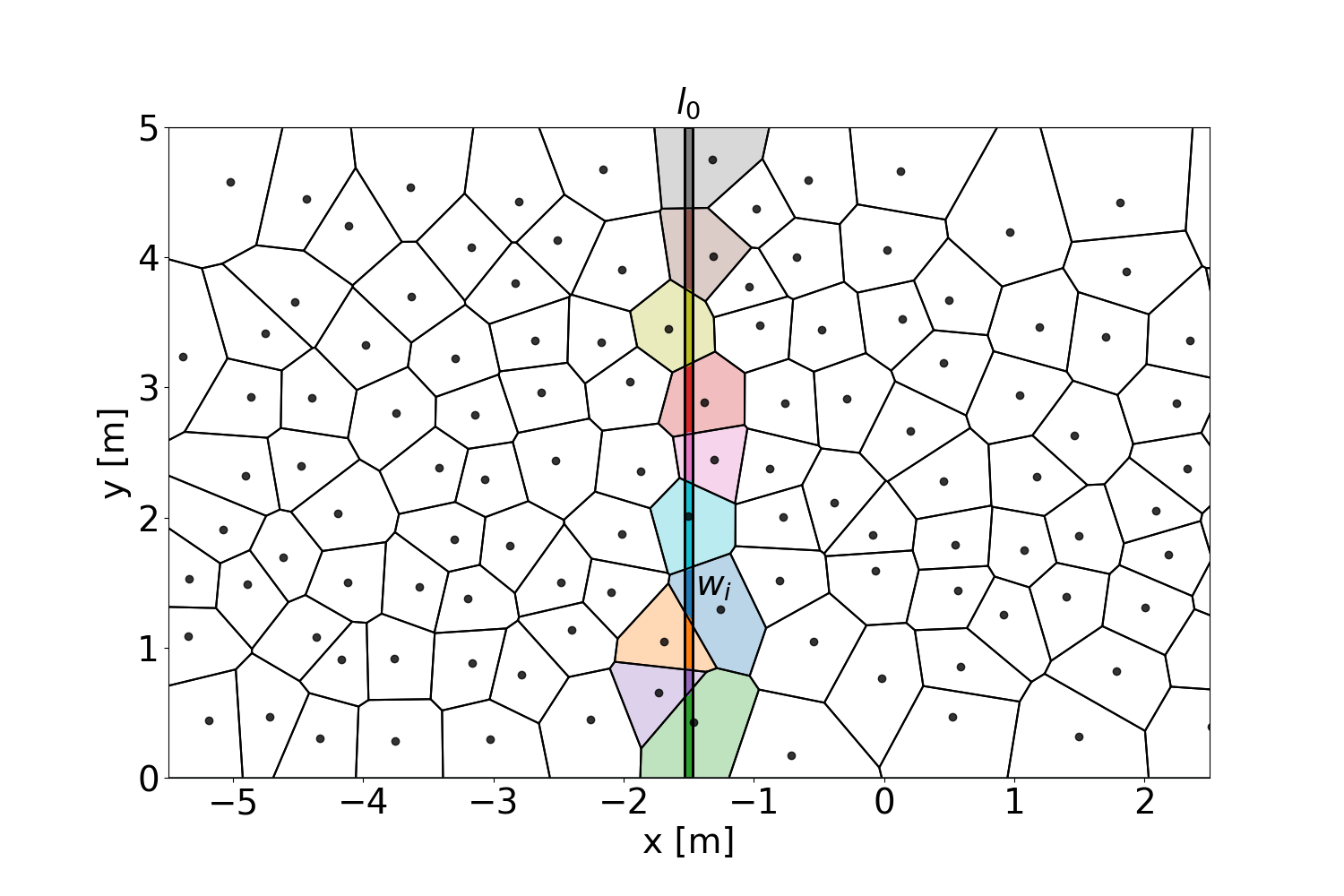}
        
    \caption{Voronoi decomposition for head positions marked by black dots. Voronoi cells intersecting with the measurement line $l_0$ are indicated by facecolors, the actual intersection $w_i$ with the line is indicated by brighter colors.}
    \label{fig:voronoi_example}
\end{figure}

All terms in the product of the sums with $i \neq j$ are zero by definition of the fields ${\boldsymbol{\rho}}(\vec{x},t)$ and $\vec{\boldsymbol{v}}(\vec{x},t)$. In addition, within a Voronoi cell $A_i(t)$, the fields are constant and thus, the integral over the measurement line $l_0$ can be replaced with the following sum

\begin{align}
J_{l_0}^{\text{v}}(t) &= \int_{y_0}^{y_1} \left( \sum_i \boldsymbol{\rho}_i(x_0,y,t) ~ \vec{\boldsymbol{v}}_i(x_0,y,t) ~ \vec{n}_{l_0}\right)  ~ \mathrm{d}y \\ &= \sum_{\{i\mid A_i(t) \in l_0 \}} \frac{\vec{v}_i(t) ~ \vec{n}_{l_0}}{A_i(t)}   ~ w_{i,l_0}(t) \\ &= \sum_{\{i\mid A_i(t) \in l_0 \}} \rho_i(t) ~ v_i(t)^{\perp_{l_0}}  ~ w_{i,l_0}(t),
\label{eq:abcd}
\end{align}

where $w_{i,l_0}(t)$ is the length of the intersection between measurement line $l_0$ and the Voronoi cell $A_i(t)$ (see \autoref{fig:voronoi_example}). This is consistent with the following definition of a flow field based on a Voronoi-decomposition:

   \begin{align}
    \vec{\boldsymbol{j}}(\vec{x},t) = \sum_i \vec{\boldsymbol{j}}_i(\vec{x},t) ~ \text{with} ~ \vec{\boldsymbol{j}}_i(\vec{x},t) &= \left\{\begin{array}{ll} \frac{\vec{v}_i(t)}{A_i(t)} : & \vec{x} \in A_i(t) \\
    0: & otherwise\end{array}\right\}.
    \end{align}

To compare the field definition by Voronoi-decomposition with classical measurements for speed and density, see Equation \autoref{eq:density_classical} and \autoref{eq:variables_<v>_x0}, we integrate over the measurement lines and calculate a mean value by dividing by the length $w$ of measurement line $l_0$.
     
\begin{equation}
v_{l_0}^{\text{v}}(t)= \frac{1}{w} \int_{y_0}^{y_1} \vec{\boldsymbol{v}}(\vec{x},t) \vec{n}_{l_0} ~ dy =  \sum_{\{i|A_i(t) \in l_0\}}^{} \vec{v}_i(t) \vec{n}_{l_0} \frac{w_{i,l_0}(t)}{w}
\label{eq:line_speed}
\end{equation}

\begin{equation}
\rho_{l_0}^{\text{v}}(t) = \frac{1}{w} \int_{y_0}^{y_1} {\boldsymbol{\rho}}(\vec{x},t) ~ dy = \sum_{\{i|A_i(t) \in l_0\}}^{} \frac{1}{A_i(t)} \frac{w_{i,l_0}(t)}{w}
\label{eq:line_density}
\end{equation}

\begin{equation}
J_{s,l_0}^{\text{v}}(t) =  \frac{1}{w} \int_{y_0}^{y_1} \vec{\boldsymbol{j}}(\vec{x},t) \vec{n}_{l_0} ~dy = \sum_{\{i|A_i(t) \in l_0\}}^{} \frac{\vec{v}_i(t)\vec{n}_{l_0}}{A_i(t)} \frac{w_{i,l_0}(t)}{w}
\label{eq:line_flow}
\end{equation}

Interestingly, a comparison of the mean density, velocity and specific flow shows that the integration of the weights results in a linear factor in each case, emphasising the relevance of the Cauchy-Schwarz inequality. 

With these fields, the flow equation for every point in time at a measurement line $l_0$ is

\begin{equation}
J_{l_0}^{\text{v}}(t) =  \int_{y_0}^{y_1} \vec{\boldsymbol{j}}(\vec{x},t) \vec{n}_{l_0} ~ dy = \sum_{\{i|A_i(t) \in l_0\}}^{} \frac{\vec{v}_i(t)\vec{n}_{l_0}}{A_i(t)} ~ w_{i,l_0}(t)  = \sum_{\{i|A_i(t) \in l_0\}}^{} \rho_i(t) ~ v_i(t)^{\perp_{l_0}} ~ w_{i,l_0}(t)
\label{eq:ghi}
\end{equation}

The following classical measurement methods based on the trajectories of particles are compared with the measurements based on the fields constructed via the Voronoi decomposition. Classical measurements are given the superscript $c$. Field measurements based on Voronoi receive the superscript $v$. For this comparison, we use data from experiments conducted under laboratory conditions. In general, video recordings with a frame rate of, e.g. 25 $\frac{\mathrm{frames}}{\mathrm{second}}$ are used. This leads to trajectories discrete in time $t_k$ with $k=1,...,M$ but continuous in space $\vec{x}_i(t_k)$. A measurement of a flow $J$ as a mean value in the time interval $\Delta t = t_N-t_0$ at the line $l_0$ according to \autoref{eq:flowequation_fields_deltat} is 

\begin{equation}
J_{l_0,\Delta t}^{\text{v}} 
= \frac{1}{M} \sum_{k=1}^{M} \left( \sum_{\{i|A_i(t_k) \in l_0\}}^{} \frac{\vec{v}_i(t_k) \Vec{n}_{l_0}}{A_i(t_k)} w_{i,l_0}(t_k)\right)
\label{eq:flowequation_fields}
\end{equation}

where $t_0=t_{k=1}$ and $t_N=t_{k=M}$. 

At high densities, several persons can walk side by side across $l_0$, and the time interval $\tau_{i,l_0}$ between two successive crossings can be rather small. When calculating the flow according to $J_{l_0,\Delta t}^{c}=\frac{N}{t_N-t_0}$, it is therefore possible that more than one pedestrian crosses $l_0$ between two consecutive frames $t_k$ and $t_{k+1}$, e.g. $t_k<t_i<t_{i+1}<t_{k+1}$. This requires an adequate interpolation of the trajectories between $t_k$ and $t_{k+1}$. The velocity of an individual at $t_k$ is calculated by

\begin{equation}
    \vec{v}_i(t_k) = \frac{\vec{x}_i(t_{k+n}) - \vec{x}_i(t_{k-n})}{t_{k+n}-t_{k-n}} 
\end{equation}

For the following measurement, we use $n=10$

\begin{figure}[htp]
    \centering
    \subfigure[]{
        \includegraphics[width=.49\textwidth]{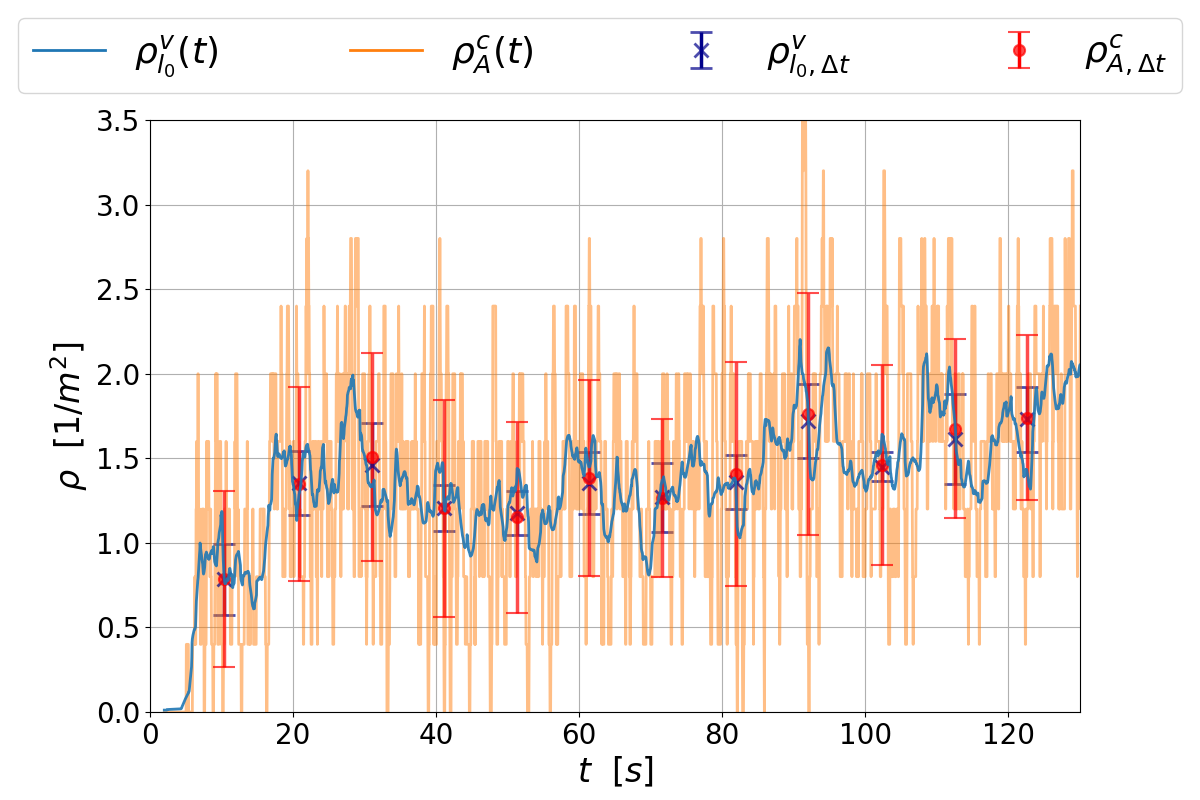}
        
    \label{fig:tseries_rho_classic_uni_03_a}}
    \hspace{-1.5em}
    \subfigure[]{
        \includegraphics[width=.49\textwidth]{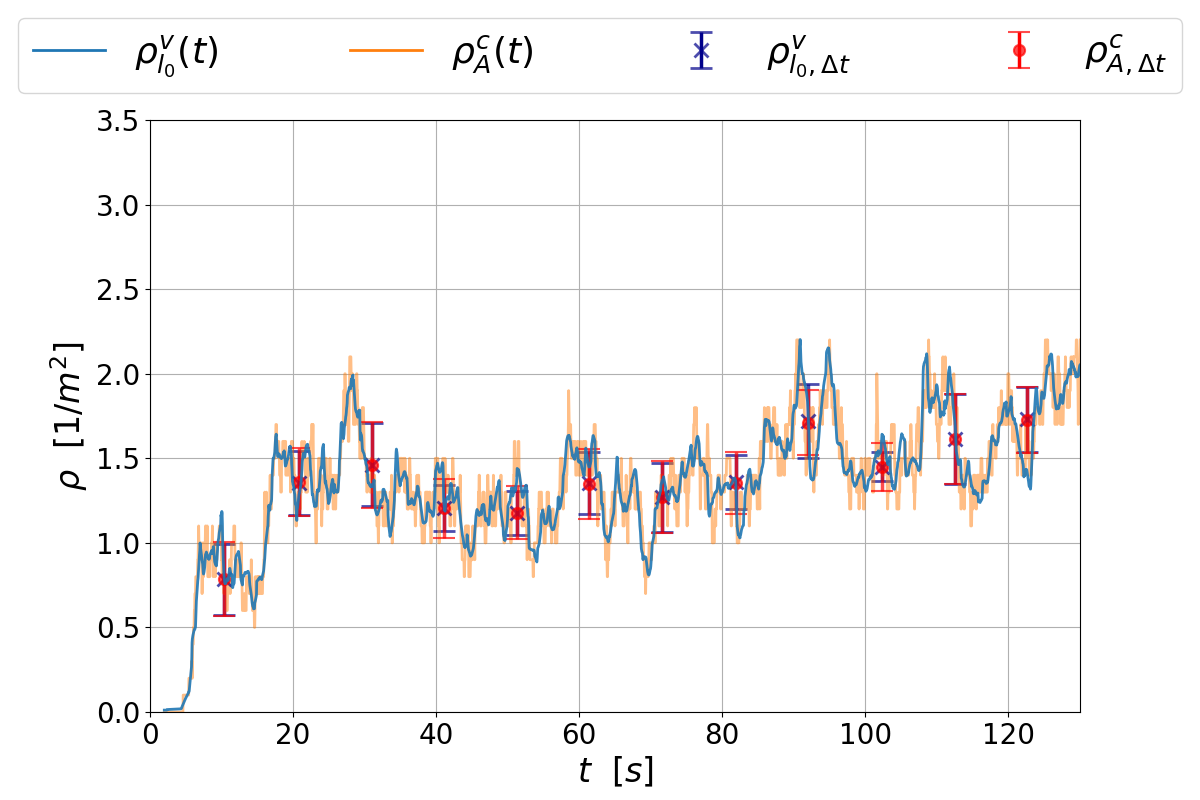}
        \label{fig:tseries_rho_classic_uni_03_b}}
    \caption{Density timeseries for experiment uni\_corr\_500\_03. Comparison between the line-density $\rho_{l_0}^{\text{v}}(t)$ (\autoref{eq:line_density}) and the classical density $\rho_A^{\text{c}}(t)$ (\autoref{eq:density_classical}). Their mean values withing time intervals $\Delta t=\SI{10}{s}$ are $\rho_{l_0,\Delta t}^{\text{v}}$ and $\rho_{A, \Delta t}^{\text{c}}$ respectively. For the classical density, the measurement area around $l_0$ has a width of a) $\Delta x=\SI{0.5}{m}$ and b) $\Delta x=\SI{2}{m}$.}
    \label{fig:tseries_rho_classic_uni_03}
\end{figure}

\begin{figure}[htp]
    \centering
    \subfigure[]{
        \includegraphics[width=.49\textwidth]{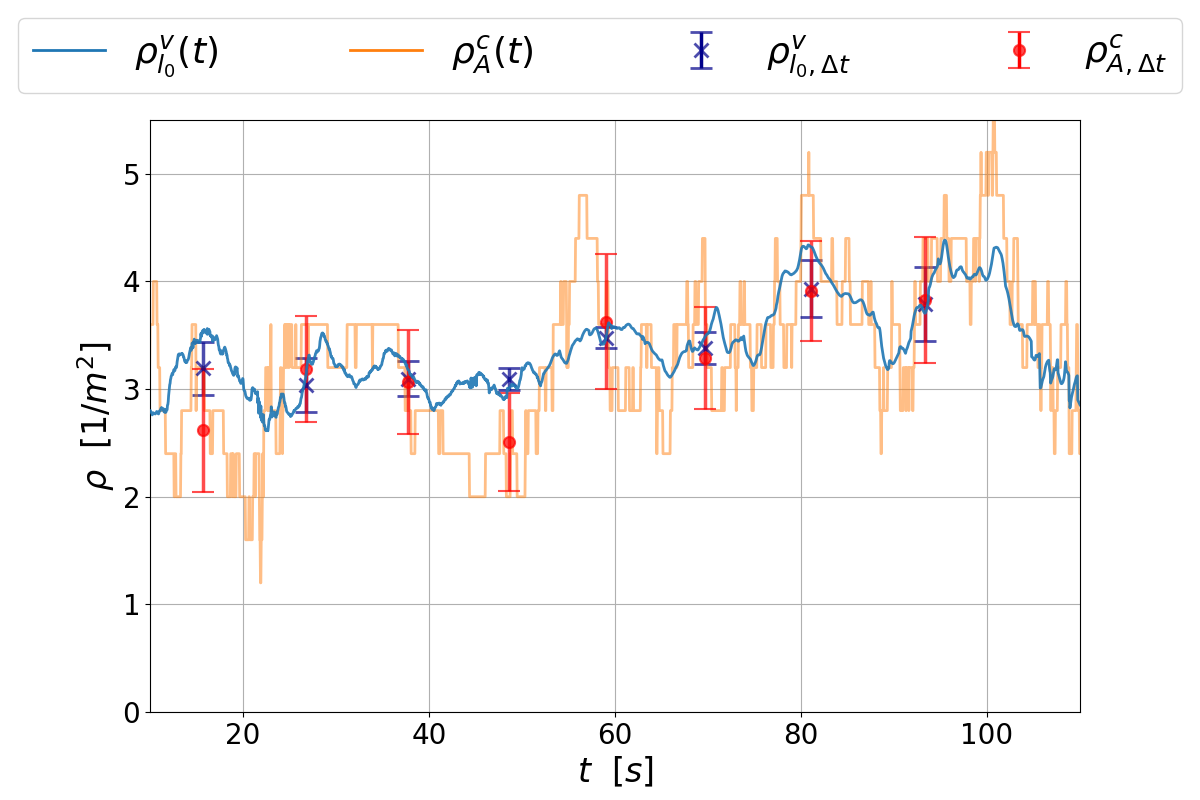}
        
    \label{fig:tseries_rho_classic_uni_10_a}}
    \hspace{-1.5em}
    \subfigure[]{
        \includegraphics[width=.49\textwidth]{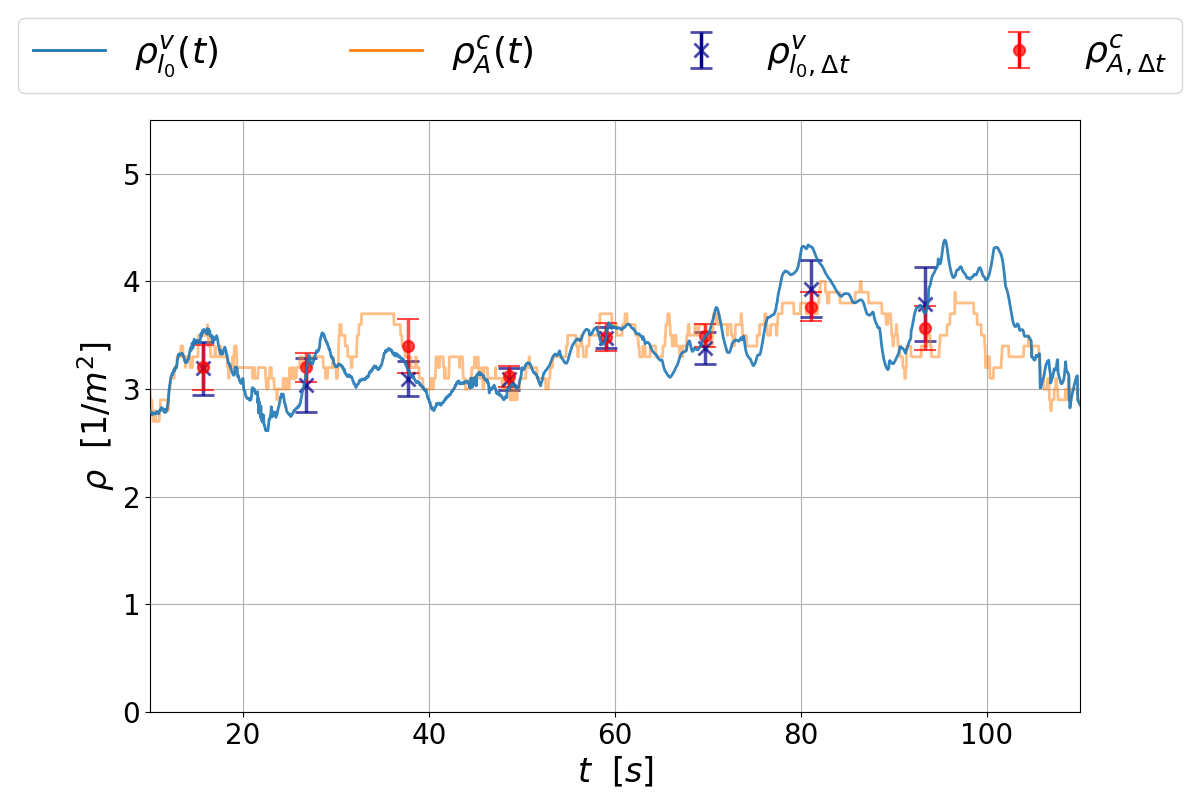}
        \label{fig:tseries_rho_classic_uni_10_b}}
    \caption{Density timeseries for experiment uni\_corr\_500\_10. Comparison between the line-density $\rho_{l_0}^{\text{v}}(t)$ (\autoref{eq:line_density}) and the classical density $\rho_A^{\text{c}}(t)$ (\autoref{eq:density_classical}). Their mean values within time intervals $\Delta t=\SI{10}{s}$ are $\rho_{l_0, \Delta t}^{\text{v}}$ and $\rho_{A, \Delta t}^{\text{c}}$ respectively. For the classical density, the measurement area around $l_0$ has a thickness of a) $\Delta x=\SI{0.5}{m}$ and b) $\Delta x = \SI{2}{m}$. }
    \label{fig:tseries_rho_classic_uni_10}
\end{figure}

As examples for unidirectional flow, we use the datasets of the \textit{uni\_corr\_500} experiments (https://doi.org/10.34735/ped.2013.6) that were acquired in a corridor with $w= \SI{5}{m}$ width in the framework of the \textit{BASIGO} project. The two subexperiments that we look at in more detail are \textit{uni\_corr\_500\_03} in which the flow is close to its maximum and \textit{uni\_corr\_500\_10}, which is the experiment with the highest density, i.e. the flow is decreased. 
In \autoref{fig:tseries_rho_classic_uni_03} and \autoref{fig:tseries_rho_classic_uni_10} we compare the line density $\rho_{l_0}^{\text{v}}(t)$ with the classical density $\rho_{A}^c(t)$ measured withing a narrow and a wide measurement area. In the case of experiment \textit{uni\_corr\_500\_03}, the mean values $\rho_{l_0,\Delta t}^{\text{v}}$ and $\rho_{A,\Delta t}^c$ within a \SI{10}{s} time interval are in good correspondence within the error limits. But for experiment \textit{uni\_corr\_500\_10}, there are some deviations between classical and field measurements of the density, indicating the relevance of spatial inhomogeneities at crowded conditions.  

\begin{figure}[htp]
    \centering
    \subfigure[]{
        \includegraphics[width=.49\textwidth]{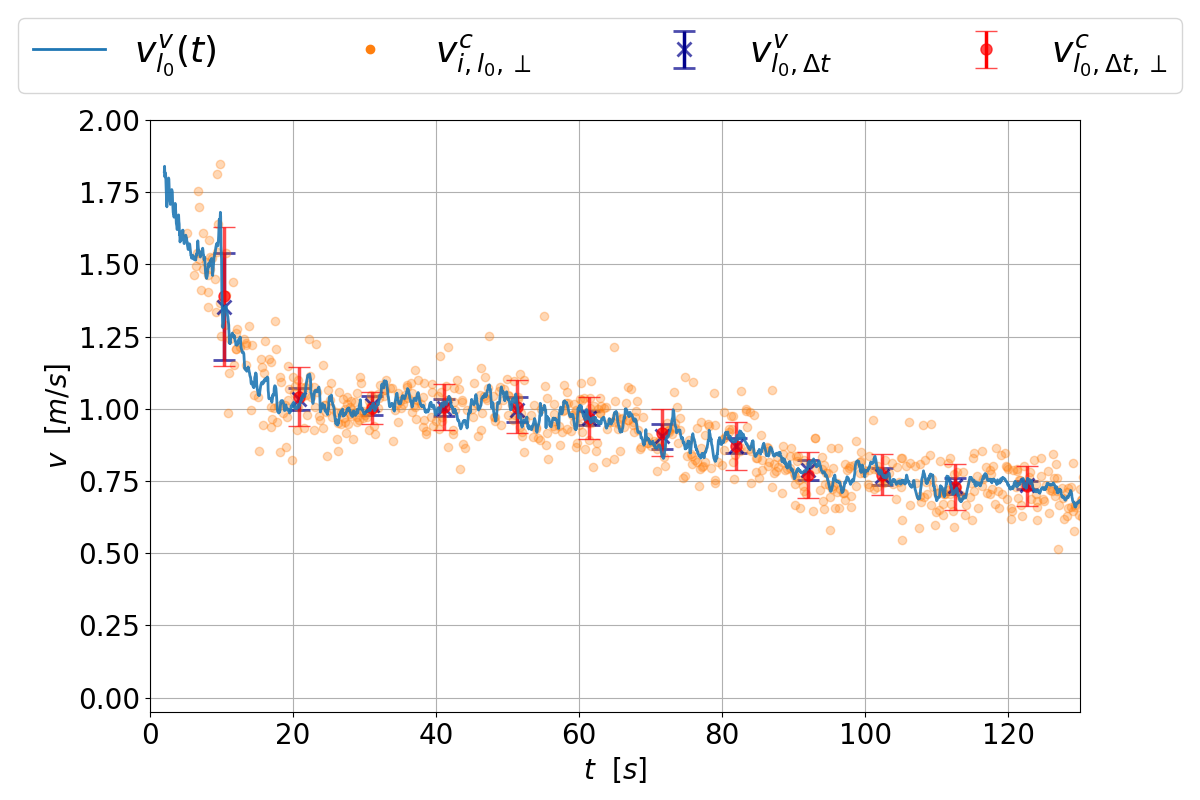}
        
    \label{fig:tseries_v_classic_uni_a}}
    \hspace{-1.5em}
    \subfigure[]{
        \includegraphics[width=.49\textwidth]{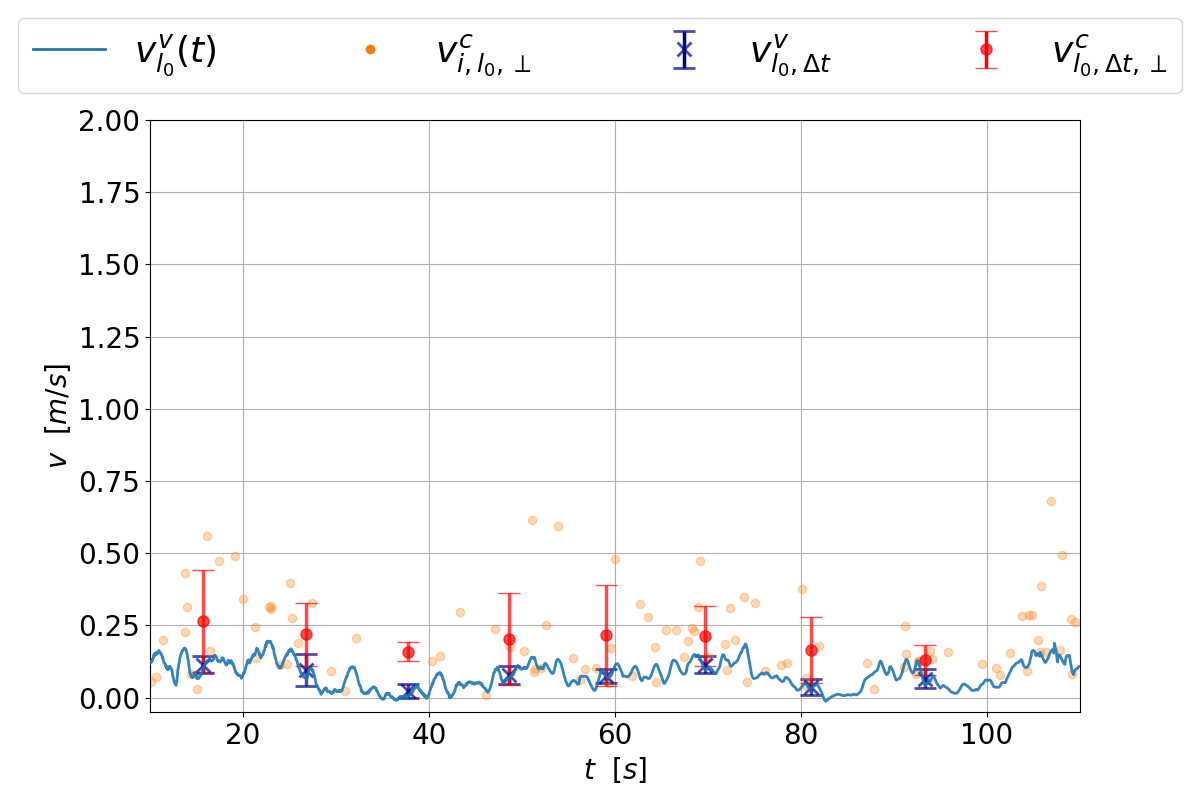}
        \label{fig:tseries_v_classic_uni_b}}
    \caption{Speed timeseries for experiments a) uni\_corr\_500\_03 and b) uni\_corr\_500\_10. Comparison between $v_{l_0}^{\text{v}}(t)$ (\autoref{eq:line_speed}) and the classical individual perpendicular crossing velocity $v_{i,l_0,\perp}^{\text{c}}$. Their mean values within time intervals $\Delta t=\SI{10}{s}$ are $v_{l_0, \Delta t}^{\text{v}}$ and $v_{l_0, \Delta t, \perp}^{\text{c}}$ (\autoref{eq:variables_<v>_x0_perp}) respectively.}
    \label{fig:tseries_v_classic_uni}
\end{figure}

A comparison between the line-speed $v_{l_0}^{\text{v}}(t)$ and the classical crossing speed $v_{i,l_0,\perp}^{\text{c}}(t)$ is shown in \autoref{fig:tseries_v_classic_uni}. Again, both methods correspond well to experiment \textit{uni\_corr\_500\_03}. But for experiment \textit{uni\_corr\_500\_10}, there are clear deviations. This can be explained as follows: When using the classical crossing speed $v_{i,l_0,\perp}^{\text{c}}(t)$, moving pedestrians are taken into account while crossing the line. Pedestrians in standstill shortly in front or behind the line are not considered. In the measurements of the line-speed based on fields $v_{l_0}^{\text{v}}(t)$, standing pedestrians whose Voronoi cell intersects the line contribute to the integral. This is exactly the difference that can be seen in Figure \autoref{fig:tseries_v_classic_uni_b}.

\begin{figure}[htp]
    \centering
    \subfigure[]{
        \includegraphics[width=.49\textwidth]{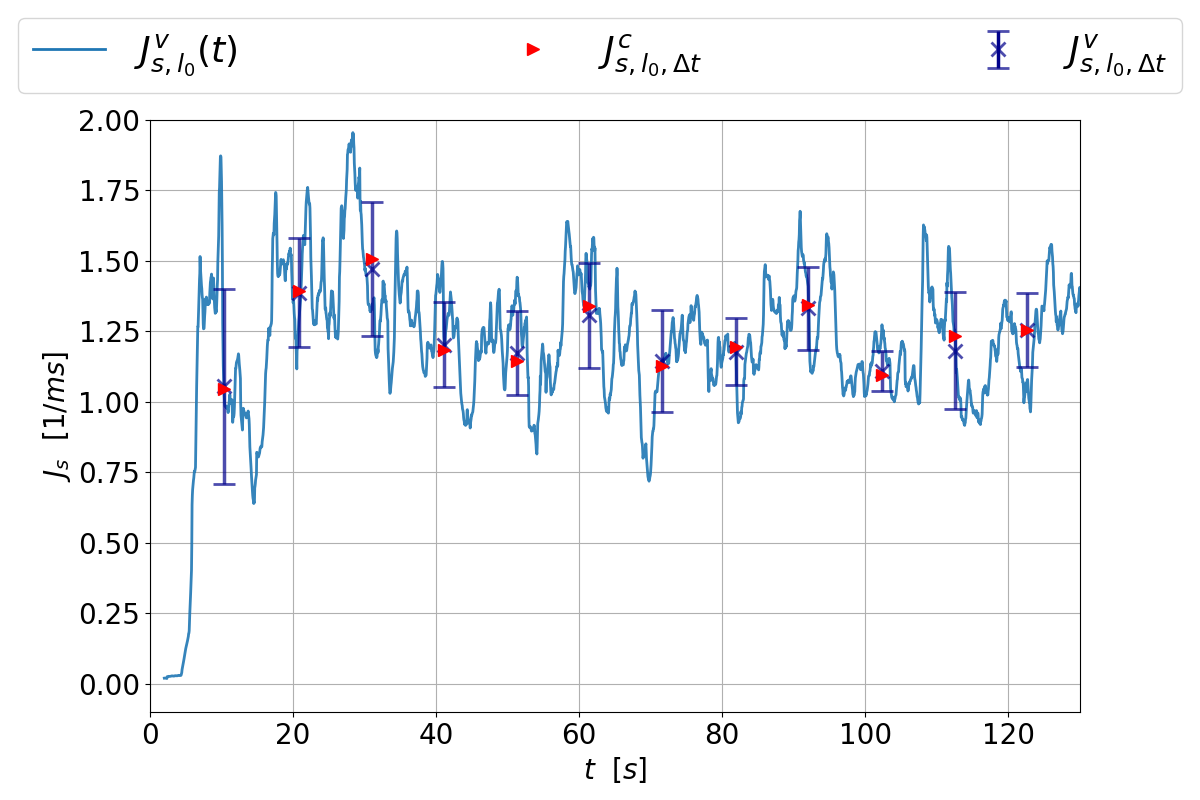}
        
    \label{fig:tseries_j_classic_uni_a}}
    \hspace{-1.5em}
    \subfigure[]{
        \includegraphics[width=.49\textwidth]{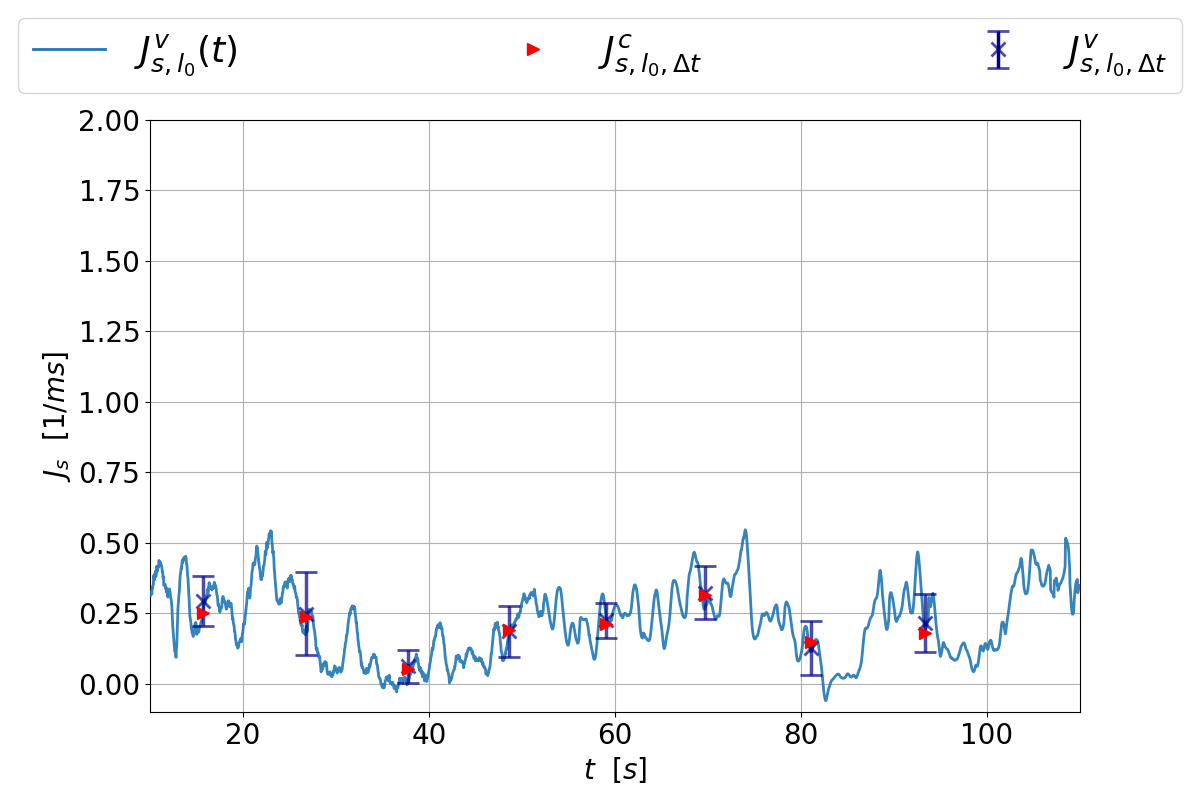}
        \label{fig:tseries_j_classic_uni_b}}
    \caption{Flow timeseries for experiments a) uni\_corr\_500\_03 and b) uni\_corr\_500\_10. Comparison between the specific flow $J_{s, l_0}^{\text{v}}(t)$ (\autoref{eq:line_flow}) with mean value $J_{s, l_0, \Delta t}^{\text{v}}$ and the classical specific flow $J_{s, l_0, \Delta t}^{\text{c}}$ measured after \autoref{eq:variables_flow_macro} with $\Delta t=\SI{10}{s}$.}
    \label{fig:tseries_j_classic_uni}
\end{figure}

Figure \autoref{fig:tseries_j_classic_uni} compares classical flow measurements according \autoref{eq:variables_flow_macro} with the new method \autoref{eq:flowequation_fields}. The comparison shows good agreement and the expected alignment with negative flow values.

\begin{figure}[htp]
    \begin{center}
        \includegraphics[width=0.75\textwidth]{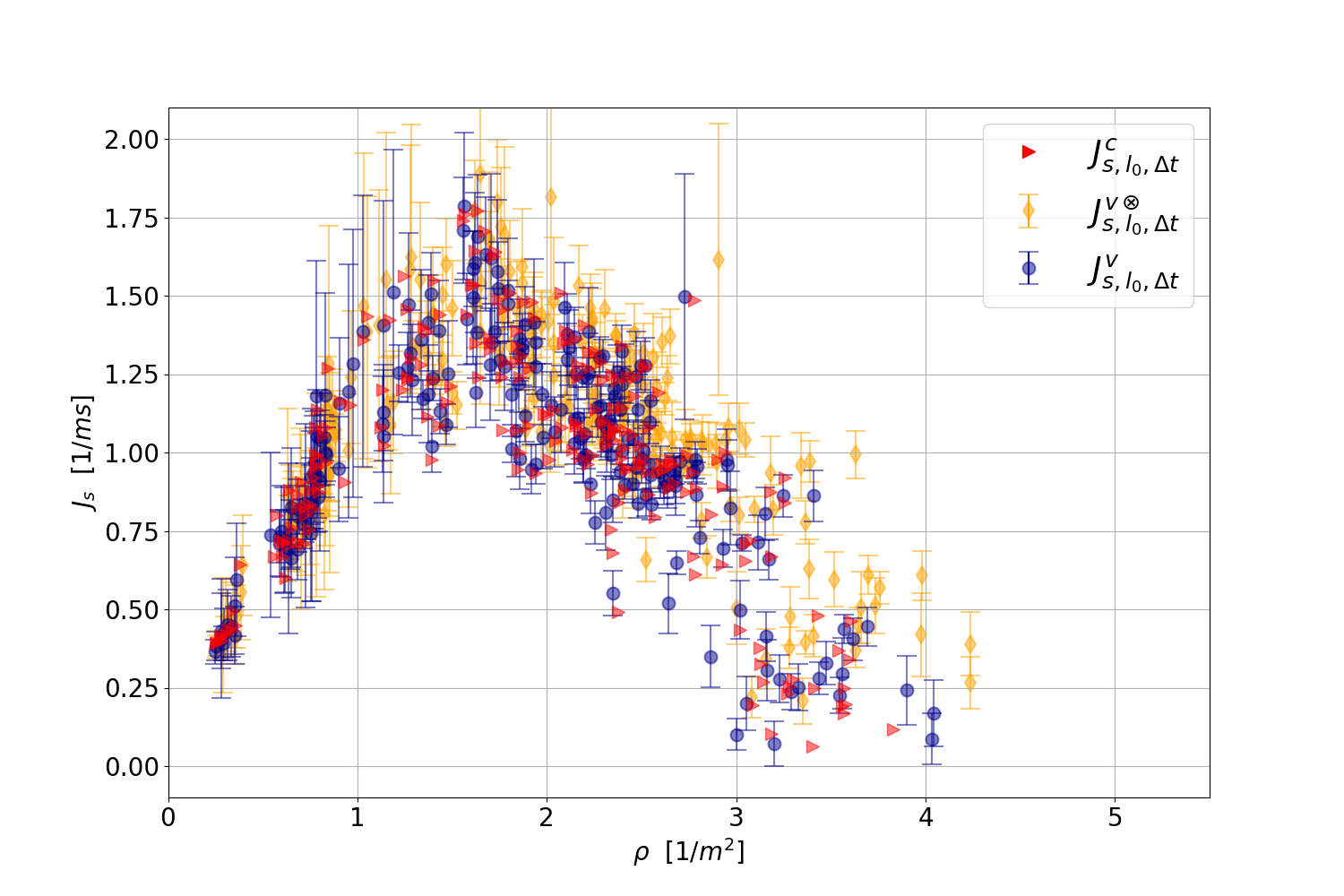}
    \end{center}
    \caption{Fundamental diagram for unidirectional flow experiments \textit{uni\_corr\_500} averaged within time intervals of $\Delta t = \SI{5}{s}$. Comparison between the new method in blue (specific flow $J_{s, l_0, \Delta t}^{\text{v}}$ (\autoref{eq:line_flow}) using line-density $\rho_{l_0, \Delta t}^{\text{v}}$ (\autoref{eq:line_density})) with classical measurements in red (specific flow $J_{s, l_0, \Delta t}^{\text{c}}$ (\autoref{eq:variables_flow_macro}) and density $\rho_{A, \Delta t}^{\text{c}}$ (\autoref{eq:density_classical}) within a measurement area with width $\Delta x = \SI{2}{m}$) and a simplified measurement in yellow (with specific flow $J_{s, l_0, \Delta t}^{\text{v}\otimes}$ (\autoref{eq:J_a}) and density $\rho_{l_0, \Delta t}^{\text{v}\otimes}$ (\autoref{eq:rho_a}).}
    \label{fig:fd_uni_mean_nt}
\end{figure}

Finally, the fundamental diagram in \autoref{fig:fd_uni_mean_nt} provides a general overview of the comparison between the proposed measurement methods of density $\rho_{l_0}^{\text{v}}$ along a line and the specific flow $J_{s, l_0, \Delta t}^{\text{v}}$ with the classical measurement methods for density $\rho_{A}^{\text{c}}$ within a measurement area and the specific flow $J_{s, l_0, \Delta t}^{\text{c}}$. The general course of the fundamental diagrams is similar. Deviations can mostly be seen in the high density regime, which results from differences in the determination of density and speed. 

Summarising, we have demonstrated a method that makes it possible to calculate all three variables, density, velocity and flow, along a line, thus avoiding mixing mean values in space and time.

\subsection{Flow equation by Voronoi fields for bi- or multidirectional streams}
\label{Bifloweq}

The equations developed in the previous subsection also ensure conformity with the continuity equation for bi- and multidirectional streams. However, we must consider the two possible main movement directions pedestrians could have when crossing a line. The calculation of the flow $J_{l_0}^{\text{v}}(t)$ uses the velocity components perpendicular to the measurement line $l_0$, including the sign to explicitly allow the perpendicular speed to be negative. This is done to account for crossings that are in the opposite direction of the desired main movement direction as a negative contribution to the measured flow. This is right for the conservation of particle numbers in the control volume but leads to an inappropriate description of the performance of the corridor cross-section, see \autoref{eq:variables_flow_macro}. To define the performance of a system for multidirectional flows at line $l_0$, a distinction must be made between pedestrians with different main movement directions. 

\begin{figure}[htp]
    \begin{center}
        \subfigure[]{
          \includegraphics[width=0.8\textwidth]{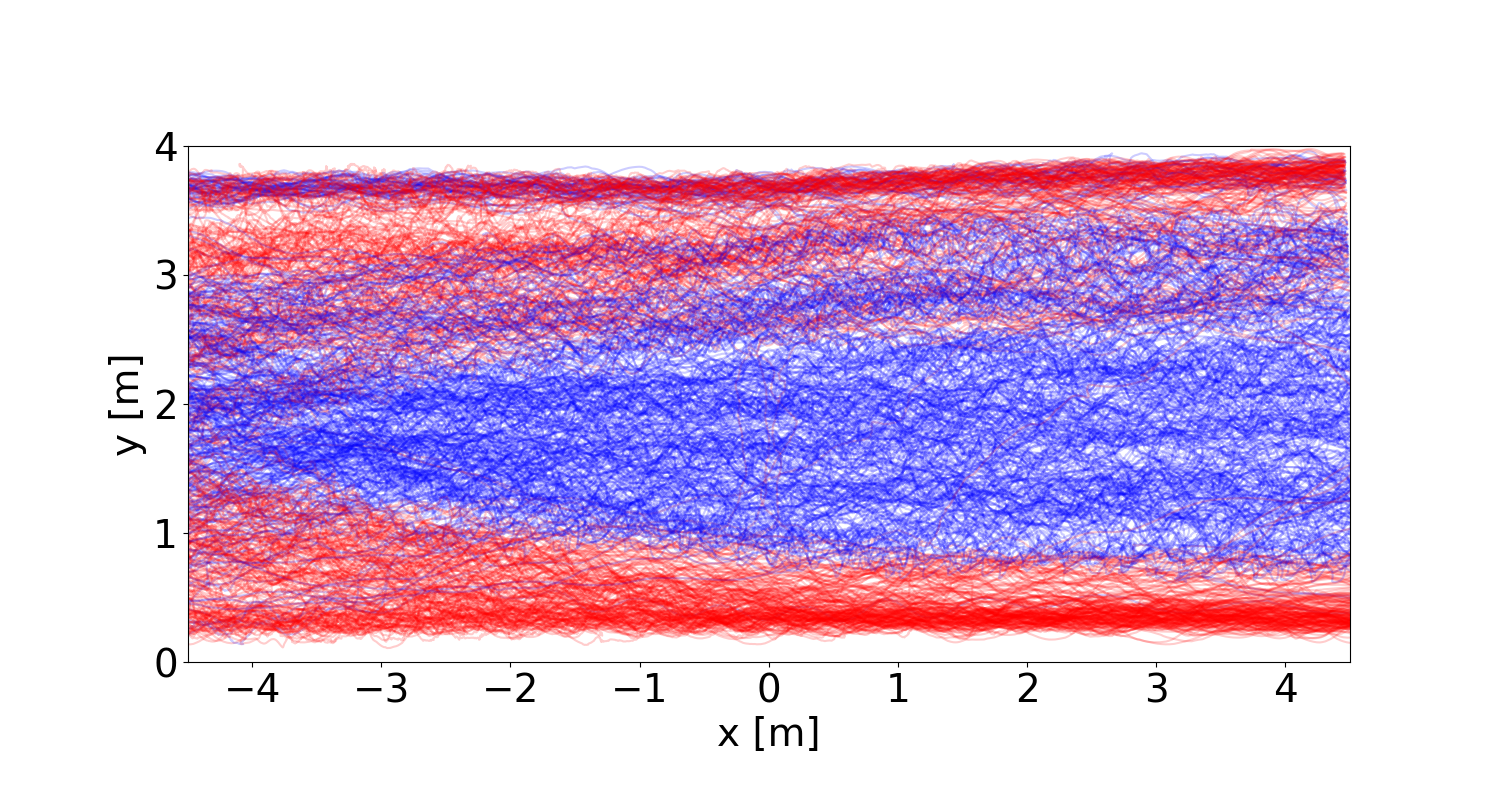}
            \label{fig:sketch_cor_spec_a}}
        \subfigure[]{
          \includegraphics[width=0.8\textwidth]{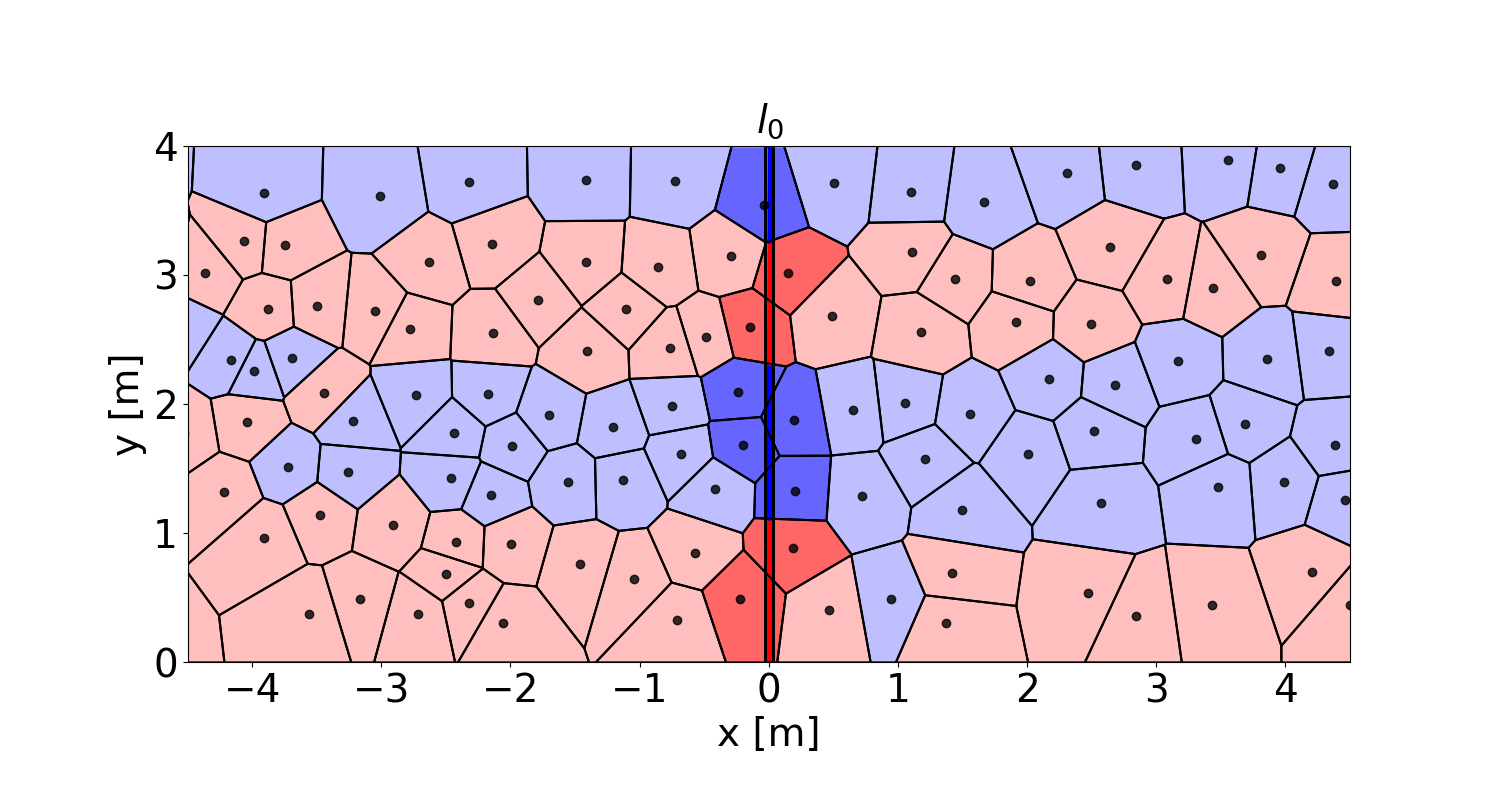}
            \label{fig:sketch_cor_spec_b}}       
    \end{center}
    \caption{Exemplary experimental dataset for bidirectional flow in a corridor. a) Individual head trajectories and b) Voronoi decomposition. Red and blue indicate the direction of the opposing main movement. In b), the Voronoi cells intersecting with measurement line $l_0$ are indicated by a brighter colour.}
    \label{fig:sketch_cor_spec}
\end{figure}

\begin{figure}[htp]
    \begin{center}
        \includegraphics[width=0.9\textwidth]{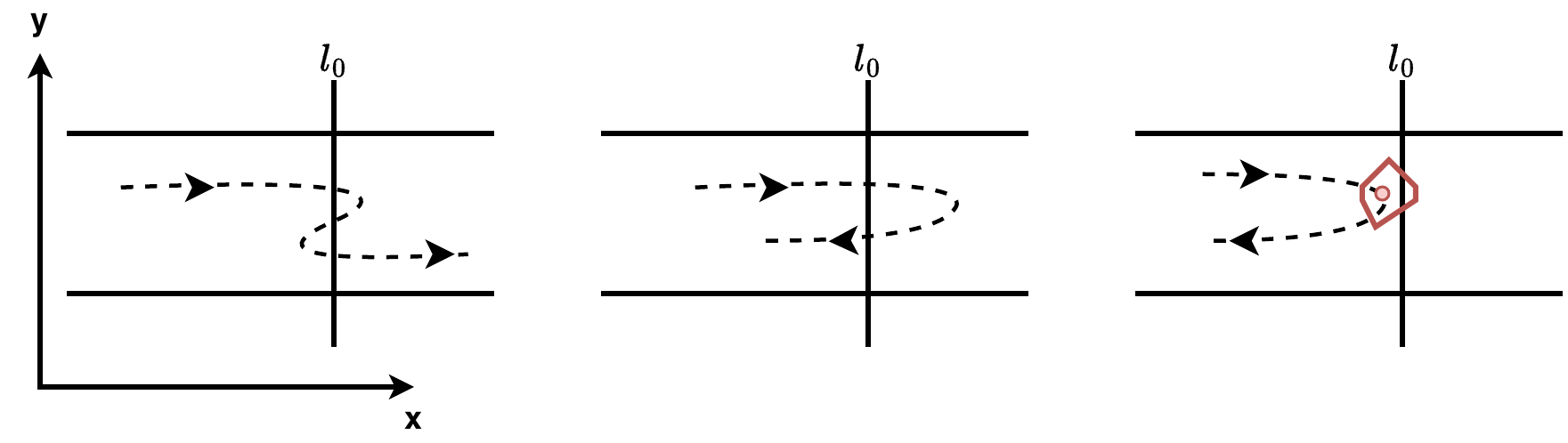}
    \end{center}
    \caption{Examples of trajectories and options for assigning trajectories to a species.}
    \label{fig:sketch_cor_spec_special}
\end{figure}

For bidirectional streams, two species, $I$ and $II$, are introduced according to the two possible main movement directions concerning the measurement line $l_0$ (see \autoref{fig:sketch_cor_spec}). For both species, the fields are defined separately. But, to still consider that these streams share the same available space, the definitions are based on a unique Voronoi decomposition considering the positions $\vec{r}_i$ of all pedestrians $i$.\\

To assign a main movement direction to each pedestrian $i$, the first crossing of the measurement line determines which species they are assigned to. In \autoref{fig:sketch_cor_spec_special}, different possibilities of crossings are sketched. In the first sketch the pedestrian has multiple crossings of the measurement line but successfully moves to the main direction of motion - here in positive $x$-direction. The figure in the middle shows an example where the pedestrian tries to move in positive $x$-direction, but does not succeed, he walks back into negative $x$-direction. The last sketch on the right shows an example where the pedestrian tries to move in positive $x$-direction, the trajectory $\Vec{r}(t)$ never crosses the measurement line $l_0$ while the Voronoi cell constructed using its position $\Vec{r}_i(t)$ crosses the line $l_0$.\\ 

The following definition ensures that every pedestrian $i$ with $A_i(t) \in l_0$ is associated with at least one species. We introduce a factor $m$ as the direction of the orthogonal velocity at the time $t_{i,l_0}=\min(t|A_i(t) \in l_0)$, when the Voronoi cell $A_i(t)$ of pedestrian $i$ touches the measurement line $l_0$ the first time by
\begin{equation}
    m = sign(\vec{n}\cdot\vec{v}(t_{i,l_0}))
\end{equation}
with $\vec{n}$ being the unit vector orthogonal to $l_0$. 

The separation of the two species, I and II, is made according to the sign of $m$:
\begin{equation}
    \text{I} = \{i |m=1\} ~ \text{and } ~  \text{II} = \{i | m=-1\}
\end{equation}
meaning that the main movement direction of all pedestrians belonging to species I is in the same direction as $\vec{n}$. In contrast, the main movement direction of all pedestrians of species II is opposite to $\vec{n}$. By this definition, all three examples shown in \autoref{fig:sketch_cor_spec_special} are assigned to species I.

The separate density field ${\boldsymbol{\rho}}^{\text{S}}$, velocity field $\boldsymbol{\vec{v}}^{\text{S}}$ and flow field $\boldsymbol{\vec{j}}^{\text{S}}$ for both species with $\text{S}=\text{I}$ and $\text{S}=\text{II}$ are 
    
   \begin{align}
    {{\boldsymbol{\rho}}^{S}}(\vec{x},t) = \sum_i {\boldsymbol{\rho}}_i(\vec{x},t) ~ \text{with} ~ {\boldsymbol{\rho}}_i(\vec{x},t) &= \left\{\begin{array}{ll} 1/|A_i(t)| : & \vec{x}\in A_i(t) \land i\in S\\
    0: & otherwise\end{array}\right\}
    \end{align}
    
   \begin{align}
    {{\boldsymbol{\vec{v}}}^{S}}(\vec{x},t) = \sum_i \boldsymbol{\vec{v}}_i(\vec{x},t) ~ \text{with} ~ \boldsymbol{\vec{v}}_i(\vec{x},t) &= \left\{\begin{array}{ll} m\cdot\vec{v}_i(t) : & \vec{x} \in A_i(t) \land i\in S \\
    0: & otherwise\end{array}\right\}
    \end{align}

   \begin{align}
    {{\boldsymbol{\vec{j}}}^{S}}(\vec{x},t) = \sum_i \boldsymbol{\vec{j}}_i(\vec{x},t) ~ \text{with} ~ \boldsymbol{\vec{j}}_i(\vec{x},t) &= \left\{\begin{array}{ll} \frac{m\cdot\vec{v}_i(t)}{A_i(t)} : & \vec{x} \in A_i(t) \land i\in S \\
    0: & otherwise\end{array}\right\}
    \end{align}

This definition allows a separate analysis of the streams in consistency with the continuity equation. To obtain the combined fields, the density and velocity fields of all pedestrians can be added. 

\begin{equation}
    \boldsymbol{\rho}(\vec{x},t) = \boldsymbol{\rho}^{\text{I}}(\vec{x},t)+\boldsymbol{\rho}^{\text{II}}(\vec{x},t)
\end{equation}

\begin{equation}
    \boldsymbol{\Vec{v}}(\vec{x},t) = \boldsymbol{\Vec{v}}^{\text{I}}(\vec{x},t)+\boldsymbol{\Vec{v}}^{\text{II}}(\vec{x},t) 
\end{equation}

Due to $\boldsymbol{\rho}^{\text{I}}(\vec{x},t) \cdot \boldsymbol{\Vec{v}}^{\text{II}}(\vec{x},t) =0$ as well as $\boldsymbol{\rho}^{\text{II}}(\vec{x},t) \cdot \boldsymbol{\Vec{v}}^{\text{I}}(\vec{x},t) =0$ also the flow fields of the two species can be added

\begin{equation}
    \boldsymbol{\Vec{j}}(\vec{x},t) = \boldsymbol{\Vec{j}}^{\text{I}}(\vec{x},t)+\boldsymbol{\Vec{j}}^{\text{II}}(\vec{x},t) 
\end{equation}

\begin{figure}[htp]
    \centering
    \subfigure[]{
        \includegraphics[width=.49\textwidth]{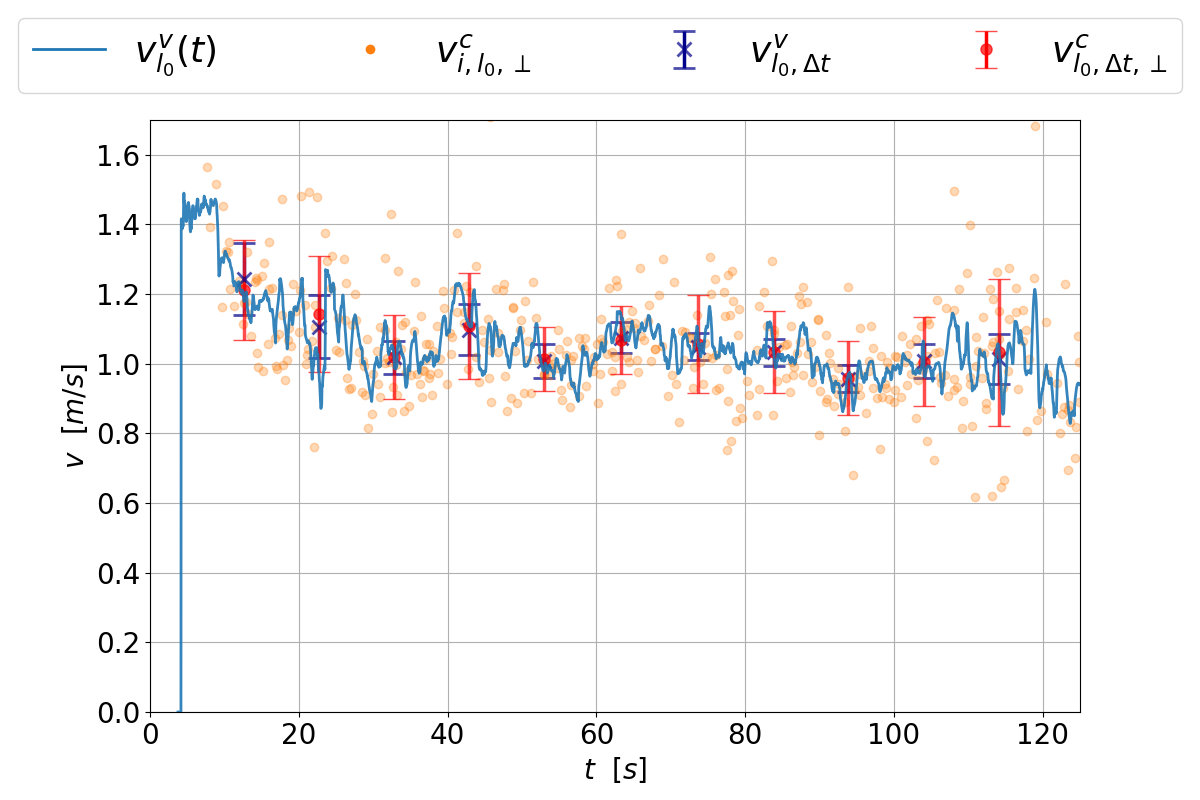}
        
    \label{fig:tseries_v_classic_bi_a}}
    \hspace{-1.5em}
    \subfigure[]{
        \includegraphics[width=.49\textwidth]{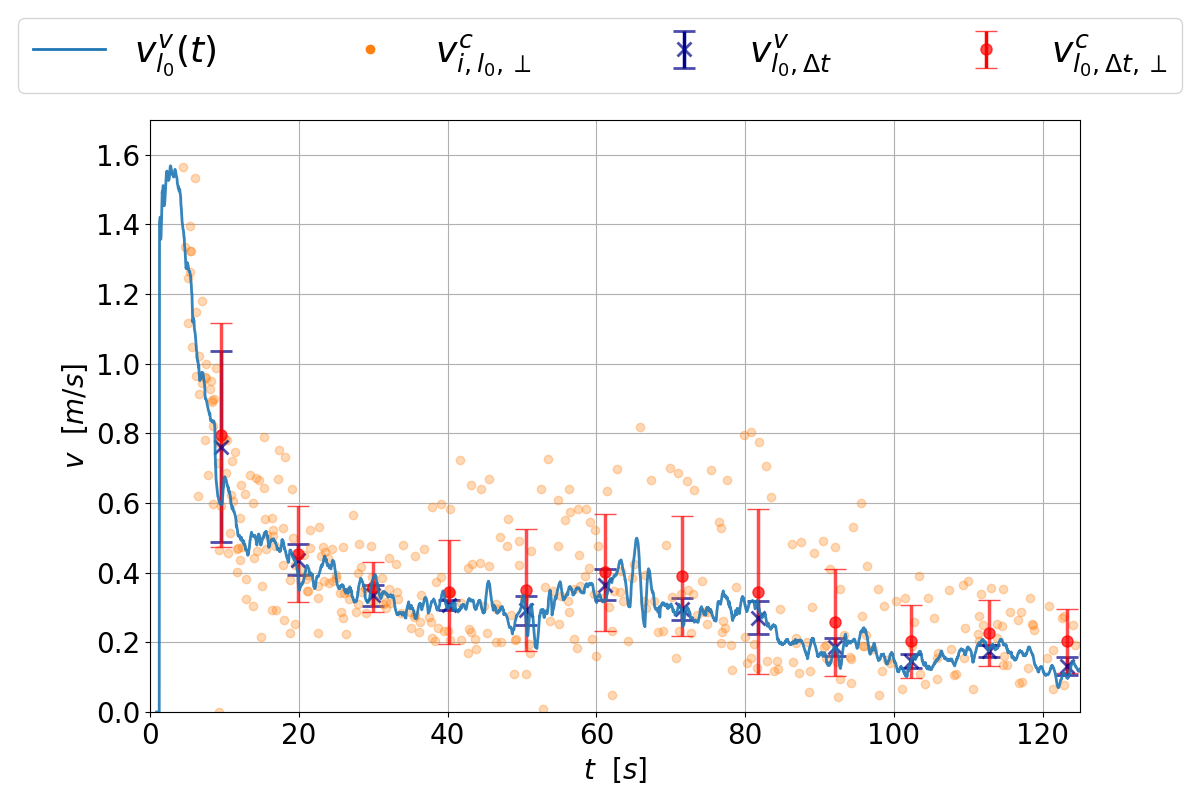}
        \label{fig:tseries_rho_classic_bi_b}}
    \caption{Speed timeseries for experiments a) bi\_corr\_400\_03 and b) bi\_corr\_400\_08. Comparison between $v_{l_0}^{\text{v}}(t)$ (\autoref{eq:line_speed}) and the classical individual perpendicular crossing velocity $v_{i,l_0,\perp}^{\text{c}}$. Their mean values within time intervals $\Delta t=\SI{10}{s}$ are $v_{l_0, \Delta t}^{\text{v}}$ and $v_{l_0, \Delta t,\perp}^{\text{c}}$ (\autoref{eq:variables_<v>_x0_perp}) respectively.}
    \label{fig:tseries_v_classic_bi}
\end{figure}

As example for bidirectional flow, the datasets of the \textit{bi\_corr\_400} experiments (https://doi.org/10.34735/ped.2013.5) with a corridor width of \SI{4}{m} are used that were acquired in the framework of the \textit{BASIGO} project. In \autoref{fig:tseries_v_classic_bi} the exemplary time series of the speed are shown. Fig. \textit{bi\_corr\_400\_03} shows the run where the flow is close to its maximum. Fig. \textit{bi\_corr\_400\_08} shows the speed of the experimental run with the highest density. The comparison between the proposed speed measurement method at the line $v_{l_0}^{\text{v}}$ with the classical vertical velocity at the time of the crossing $v_{i,l_0,\perp}^{\text{c}}$ is similar to that of the unidirectional experiments. The velocity values of the experiment in which the flow is close to maximum are in good correspondence. However, some deviations are visible for the experiment with the highest density values in which the flow is small. 

\begin{figure}[htp]
    \begin{center}
        \includegraphics[width=0.75\textwidth]{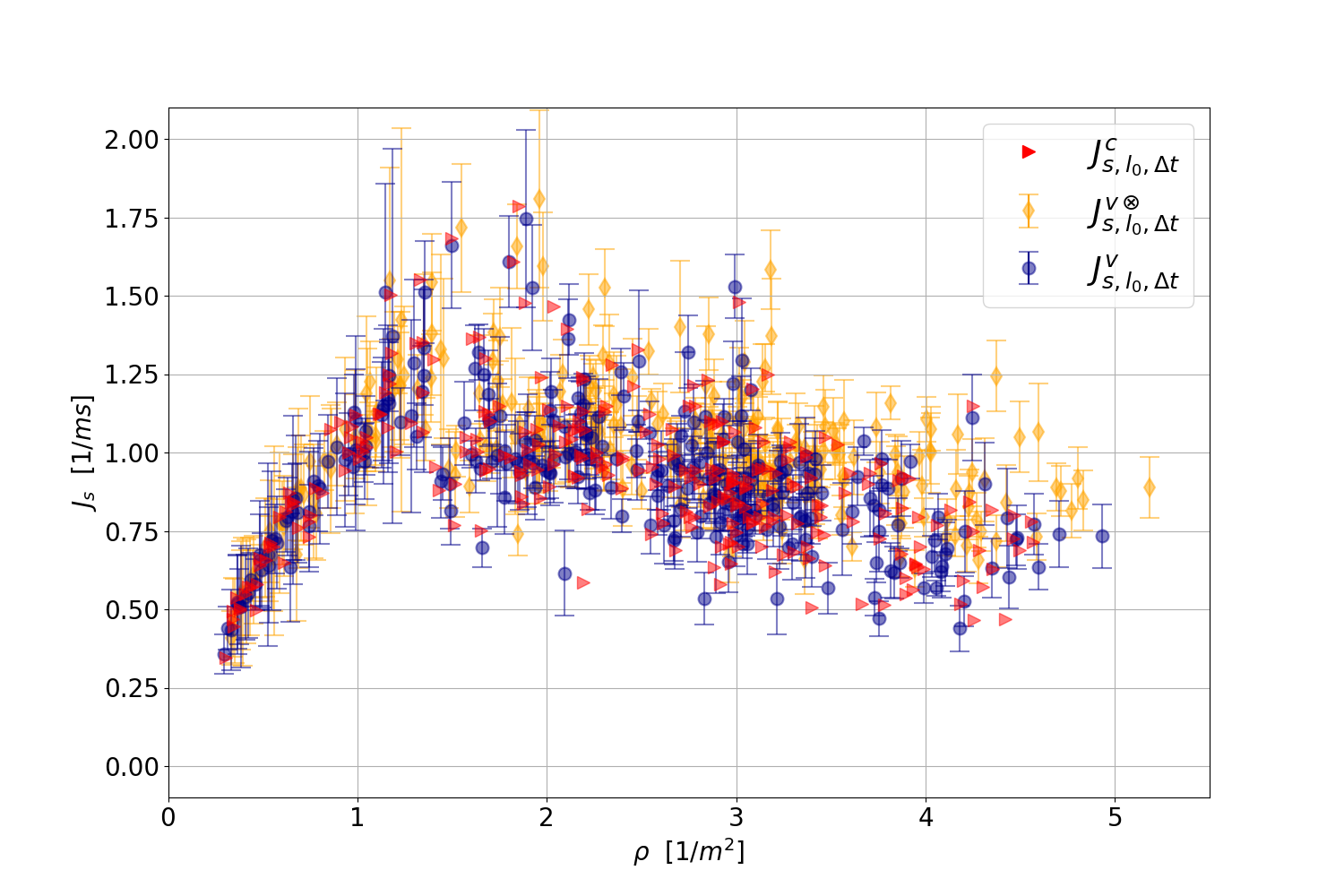}
    \end{center}
    \caption{ Fundamental diagram for bidirectional flow experiments \textit{bi\_corr\_400} averaged within time intervals of $\Delta t = \SI{5}{s}$. Comparison between the new method in blue (specific flow $J_{s, l_0, \Delta t}^{\text{v}}$ (\autoref{eq:line_flow}) using line-density $\rho_{l_0, \Delta t}^{\text{v}}$ (\autoref{eq:line_density})) with classical measurements in red (specific flow $J_{s, l_0, \Delta t}^{\text{c}}$ (\autoref{eq:variables_flow_macro}) and density $\rho_{A, \Delta t}^{\text{c}}$ (\autoref{eq:density_classical}) within a measurement area with width $\Delta x = \SI{2}{m}$) and a simplified measurement in yellow (with specific flow $J_{s, l_0, \Delta t}^{\text{v}\otimes}$ (\autoref{eq:J_a}) and density $\rho_{l_0, \Delta t}^{\text{v}\otimes}$ (\autoref{eq:rho_a}).}
    \label{fig:fd_bi_mean_nt}
\end{figure}

Accordingly, the general course of the fundamental diagram (see \autoref{fig:fd_bi_mean_nt}) calculated at the line with the proposed method is in good correspondence with the classical discrete measurements at the line.  
 
\section{Simplifications of the flow equation}

In principle, no measurement specification or definition of velocity, density or flow commonly used in the literature can be called incorrect. Some include more information, others less so. Some show large, some small fluctuations. Some emphasise boundary effects, others less, etc. In this chapter, different variants to calculate the speed, the density and the specific flow at measurement line $l_0$ are introduced and analysed. These are based on definitions that were and are used in the literature and could be regarded as simplifications of the definitions introduced in the previous section.
Simplification here means a less unambiguous consideration of information about variables that describe the movement of pedestrians in relation to a measurement line. An example would be the magnitude of a velocity vector which leaves an infinite number of options open with regard to the direction of the movement in relation to the measuring line.

The comparison starts with the strongest simplification and refined step by step to more agreement with a definition in conformance to the continuity equation.\\ 
The comparison starts with the speed. First, simply the norm $\|\vec{v}_i(t)\|$ of the individual velocities is considered and the standard mean value in the form $\bar{x} = \frac{1}{N} \sum_{i=1}^{N} x_i $ is calculated. The resulting speed is denoted as $v_{l_0}^{\text{v}\otimes}(t)$: 

\begin{equation}
v_{l_0}^{\text{v}\otimes}(t) = \frac{1}{N_{l_0}(t)} \sum_{\{i|A_i(t) \in l_0\}}^{} \|\vec{v}_i(t)\|
\label{eq:v_a}
\end{equation}

Instead of the standard mean value, the weighting term $\frac{w_i(t)}{w}$ is introduced to represent the integral along the measurement line $l_0$ and denote the resulting speed as $v_{l_0}^{\text{v}\dagger}(t)$:

\begin{equation}
v_{l_0}^{\text{v}\dagger}(t) =  \sum_{\{i|A_i(t) \in l_0\}}^{} \|\vec{v}_i(t)\| \frac{w_i(t)}{w}
\label{eq:v_b}
\end{equation}

In the next step, only the norm of normal component $\|\vec{v}_i(t)\vec{n}\|$ of the velocity is taken into account. The resulting speed is denoted as $v_{l_0}^{\text{v}*}(t)$:

\begin{equation}
v_{l_0}^{\text{v}*}(t) =  \sum_{\{i|A_i(t) \in l_0\}}^{} \|\vec{v}_i(t) \Vec{n}_{l_0}\| \frac{w_i(t)}{w}
\label{eq:v_c}
\end{equation}

As a last step, the norm is omitted to consider the negative contributions of the individual velocities and the resulting orthogonal speed is denoted as $v_{l_0}^{\text{v}}(t)$:

\begin{equation}
v_{l_0}^{\text{v}}(t) =  \sum_{\{i|A_i(t) \in l_0\}}^{} \vec{v}_i(t) \Vec{n}_{l_0} \frac{w_i(t)}{w}
\label{eq:v_d}
\end{equation}
The definition of $v_{l_0}^{\text{v}}(t)$ is equivalent to the definition in \autoref{eq:line_speed}, which conforms with the continuity equation. 

The mean speed $v_{l_0,\Delta t}^{\text{v}}$ within time interval $\Delta t$ can then be obtained by
\begin{equation}
v_{l_0,\Delta t}^{\text{v}} = \frac{1}{M_{\Delta t}}\sum_{k|t_k \in \Delta t}v_{l_0}^{\text{v}}(t_k)   
\label{eq:v_mean_delt}
\end{equation}
in which $v_{l_0}^{\text{v}}(t_k)$ can be the speed according to equations \autoref{eq:v_a}, \autoref{eq:v_b}, \autoref{eq:v_c} or \autoref{eq:v_d}.\\

For the calculation of the line density $\rho_{l_0}^{\text{v}}(t)$, two strategies could be compared. In the simplified case, the standard mean value of all individual Voronoi densities $\rho_i^{\text{v}}(t)$ are used. The resulting line density is denoted $\rho_{l_0}^{\text{v}\otimes}(t)$:

\begin{equation}
\rho_{l_0}^{\text{v}\otimes}(t) = \frac{1}{N_{l_0}(t)} \sum_{\{i|A_i(t) \in l_0\}}^{} \rho_i^{\text{v}}(t).
\label{eq:rho_a}
\end{equation}

Introducing the weighting term $\frac{w_i(t)}{w}$ to represent the line integral results in the density $\rho_{l_0}^{\text{v}}(t)$, which is following the definition of the line density in \autoref{eq:line_density}:

\begin{equation}
\rho_{l_0}^{\text{v}}(t) =  \sum_{\{i|A_i(t) \in l_0\}}^{} \rho_i^{\text{v}}(t) \frac{w_i(t)}{w}.
\label{eq:rho_b}
\end{equation}

Again, the mean density within time interval $\Delta t$ is given by 
\begin{equation}
\rho_{l_0,\Delta t}^{\text{v}} = \frac{1}{M_{\Delta t}}\sum_{k|t_k \in \Delta t}\rho_{l_0}^{\text{v}}(t_k), 
\label{eq:rho_mean_delt}
\end{equation}
with $\rho_{l_0}^{\text{v}}(t_k)$ being the line density at time $t_k$ at $l_0$ calculated by equations \autoref{eq:rho_a} or \autoref{eq:rho_b}.

Following the strategies to calculate the speed and the density, we now consider the same simplifications for the flow.\\
Again, only the norm $\|\vec{v}_i(t)\|$ of the individual Voronoi velocities is considered and the standard mean value for line $l_0$ is calculated. The resulting specific flow is denoted as $J_{s,l_0}^{\text{v}\otimes}(t)$:

\begin{equation}
J_{s,l_0}^{\text{v}\otimes}(t) = \frac{1}{N_{l_0(t)}}\sum_{\{i|A_i(t) \in l_0\}}^{} \frac{\|\vec{v}_i(t)\|}{A_i(t)}   
\label{eq:J_a}
\end{equation}

In the second step, the weighting term $\frac{w_i(t)}{w}$ is used to represent the line integral and denote the resulting specific flow as $J_{s,l_0}^{\text{v}\dagger}(t)$:

\begin{equation}
J_{s,l_0}^{\text{v}\dagger}(t) = \sum_{\{i|A_i(t) \in l_0\}}^{} \frac{\|\vec{v}_i(t)\|}{A_i(t)} \frac{w_i}{w} 
\label{eq:J_b}
\end{equation}

Then, only the norms $\|\vec{v_i}(t)\vec{n}\|$ of the normal components of the individual Voronoi velocities resulting in the specific flow $J_{s,l_0}^{\text{v}\ast}(t)$ are considered as

\begin{equation}
J_{s,l_0}^{\text{v}\ast}(t) = \sum_{\{i|A_i(t) \in l_0\}}^{} \frac{\|\vec{v}_i(t) \Vec{n}\|}{A_i(t)} \frac{w_i}{w}  
\label{eq:J_c}
\end{equation}

Finally, the individual normal velocities are used as $\vec{v}_i(t)\vec{n}$ to account negative contributions too. The resulting specific flow is denoted as $J_{s,l_0}^{\text{v}}(t)$:

\begin{equation}
J_{s,l_0}^{\text{v}} (t) = \sum_{\{i|A_i(t) \in l_0\}}^{} \frac{\vec{v}_i(t) \Vec{n}}{A_i(t)} \frac{w_i}{w},  
\label{eq:J_d}
\end{equation}

which conforms with the continuity equation definition of the flow in \autoref{eq:line_flow}.%
\\

The mean specific flow within time interval $\Delta t$ is
\begin{equation}
J_{s,l_0,\Delta t}^{\text{v}} = \frac{1}{M_{\Delta t}}\sum_{k|t_k \in \Delta t}J_{l_0}^v(t_k), 
\label{eq:j_mean_delt}
\end{equation}
with $J_{s,l_0}^{\text{v}}(t_k)$ being the flow at time $t_k$ at $l_0$ calculated by equations  \autoref{eq:J_a}, \autoref{eq:J_b}, \autoref{eq:J_c} or \autoref{eq:J_d}.%
\\

Up to here, the mean value of the product of speed and density $\overline{(\rho\cdot u)}_{l_0}^{\text{v}}$ has been used for the calculation of the local mean value of the specific flow $\overline{J}_{s,l_0}^{\text{v}}$ as in
\begin{equation}
    \overline{J}_{s,l_0}^{\text{v}}(t_k) = \overline{(\rho\cdot u)}_{l_0}^{\text{v}}(t_k).
\label{eq:j_average}
\end{equation}    
If instead, the product of the individual mean values $\overline{\rho}_{l_0}^{\text{v}}$ and $\overline{u}_{l_0}^{\text{v}}$ is used, as in
\begin{equation}
    \overline{J}_{s,l_0}^{\text{v}}(t_k) = \overline{\rho}_{l_0}^{\text{v}}(t_k) \cdot \overline{u}_{l_0}^{\text{v}}(t_k)
\label{eq:j_average2}    
\end{equation}
then, according to the Cauchy-Schwartz inequality (\autoref{eq:cauchy}), the resulting value for $\overline{J}_{s,l_0}^{\text{v}}$ is only consistent with that in \autoref{eq:j_average} if velocity and density are linearly dependent. This might be the case in the free-flow branch of the fundamental diagram, but but even here there are measurements that cast doubt on this. In addition it is  whether a linear dependence could be assumed in the congested branch (higher density). 
Applying this simplification (\autoref{eq:j_average2}) on $J_{s,l_0}^{\text{v}}$ and $J_{s,l_0}^{\text{v}\otimes}$ yields
\begin{equation}
 J_{s,l_0}^{\text{v}'} (t) = \left[ \sum_{\{i|A_i(t) \in l_0\}}^{} \vec{v}_i(t) \Vec{n} \frac{w_i}{w} \right] \cdot \left[ \sum_{\{i|A_i(t) \in l_0\}}^{} \frac{1}{A_i(t)} \frac{w_i}{w} \right]
\label{eq:J_d_star}
\end{equation}
and
\begin{equation}
J_{s,l_0}^{\text{v}\otimes'} (t) = \left[ \frac{1}{N_{l_0(t)}}\sum_{\{i|A_i(t) \in l_0\}}^{} \|\vec{v}_i(t)\|\right] \cdot \left[ \frac{1}{N_{l_0(t)}}\sum_{\{i|A_i(t) \in l_0\}}^{} \frac{1}{A_i(t)} \right]   
\label{eq:J_a_star}
\end{equation}

\begin{table}
    \centering
    \begin{tabular}{l|rrrrr}
        \textbf{Experiment} &$J_{s,l_0}^{\text{v}\otimes'}$ &$J_{s,l_0}^{\text{v}\otimes}$ &$J_{s,l_0}^{\text{v}\dagger}$ &$J_{s,l_0}^{\text{v}\ast}$ &$J_{s,l_0}^{\text{v}}$\\
        \hline
        uni\_corr\_03  &6.5\%  &6.0\%  &1.9\%  &1.9\%  &1.9\%\\
        uni\_corr\_10  &143.1\%  &122.2\%  &117.7\%  &45.0\%  &12.3\% \\
        bi\_corr\_03 &7.6\%  &6.6\%  &1.3\%  &1.1\%  &1.1\% \\
        bi\_corr\_08 &33.3\%  &23.3\%  &23.4\%  &4.0\%  &3.4\% \\
    \end{tabular}
    \caption{Root mean square (RMS) deviation (\autoref{eq:rms}) between the time averages (\autoref{eq:j_mean_delt}) calculated for the different simplifications of the specific flow and the classical specific flow $J_{s,l0 ,\Delta t}^{\text{c}}$ (\autoref{eq:variables_flow_macro}) measured at the same time intervals for different experimental datasets. With an increasing degree of simplification the RMS value increases.}
    \label{tab:J_simplifactions_rms}
\end{table}

\begin{figure}[h!]
    \includegraphics[width=0.485\textwidth]{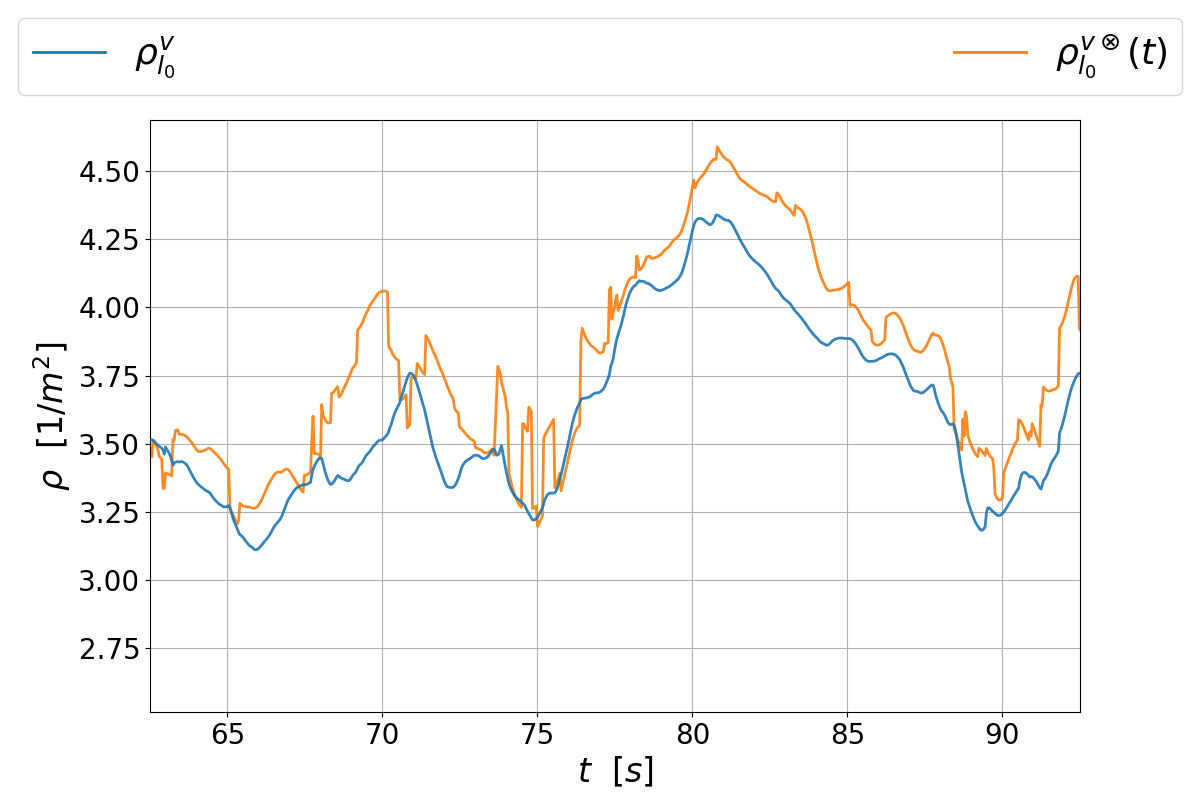}
    \includegraphics[width=0.485\textwidth]{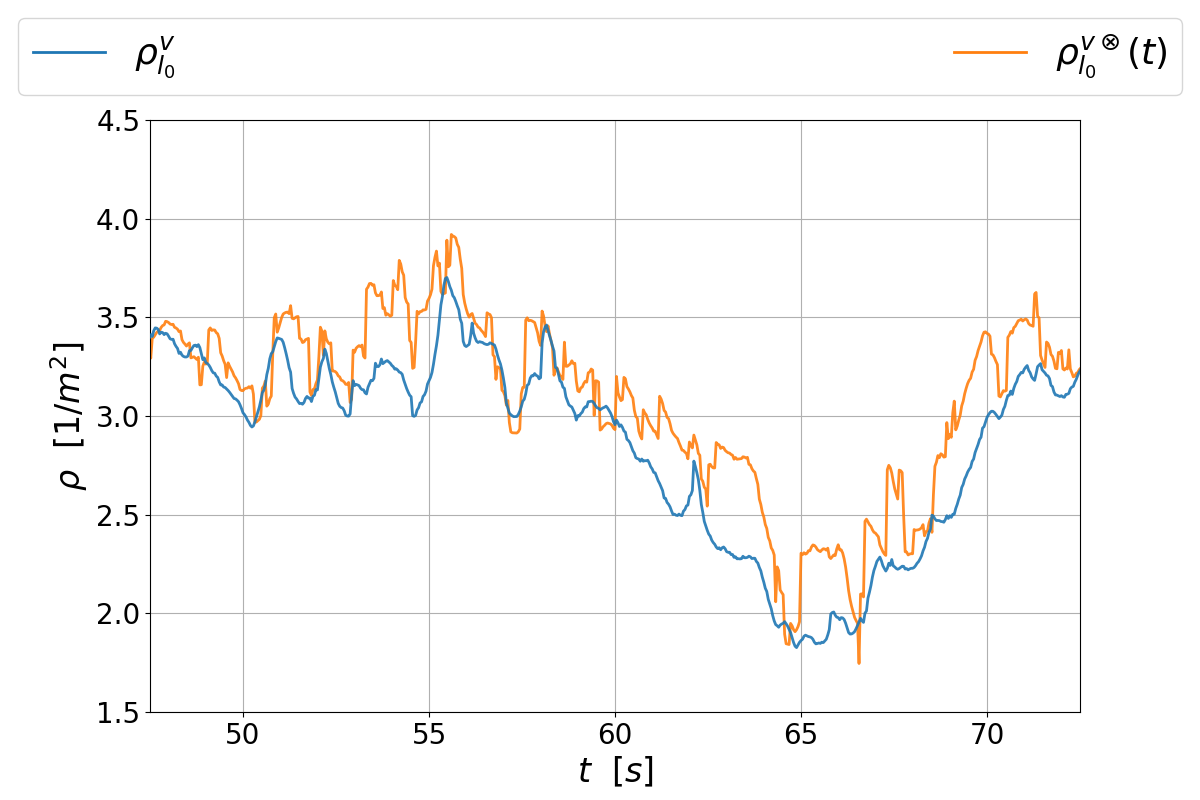}\\
    \includegraphics[width=0.485\textwidth]{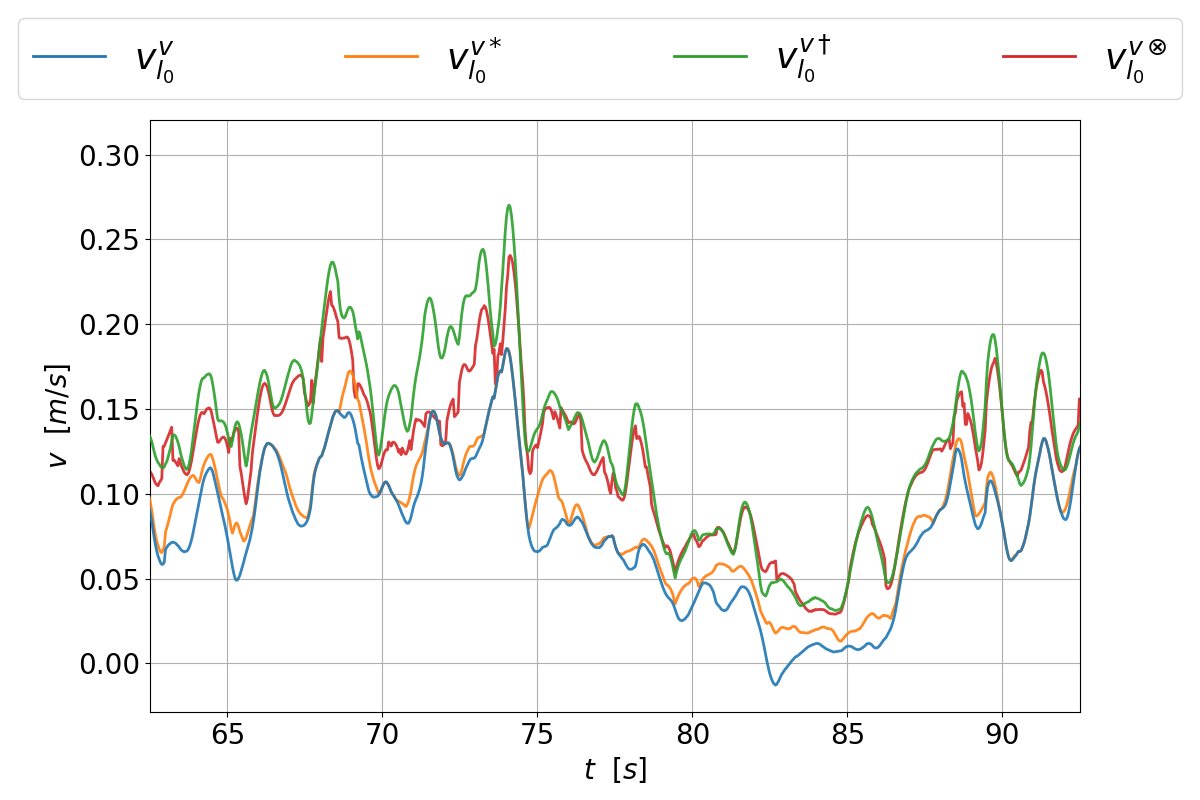}
    \includegraphics[width=0.485\textwidth]{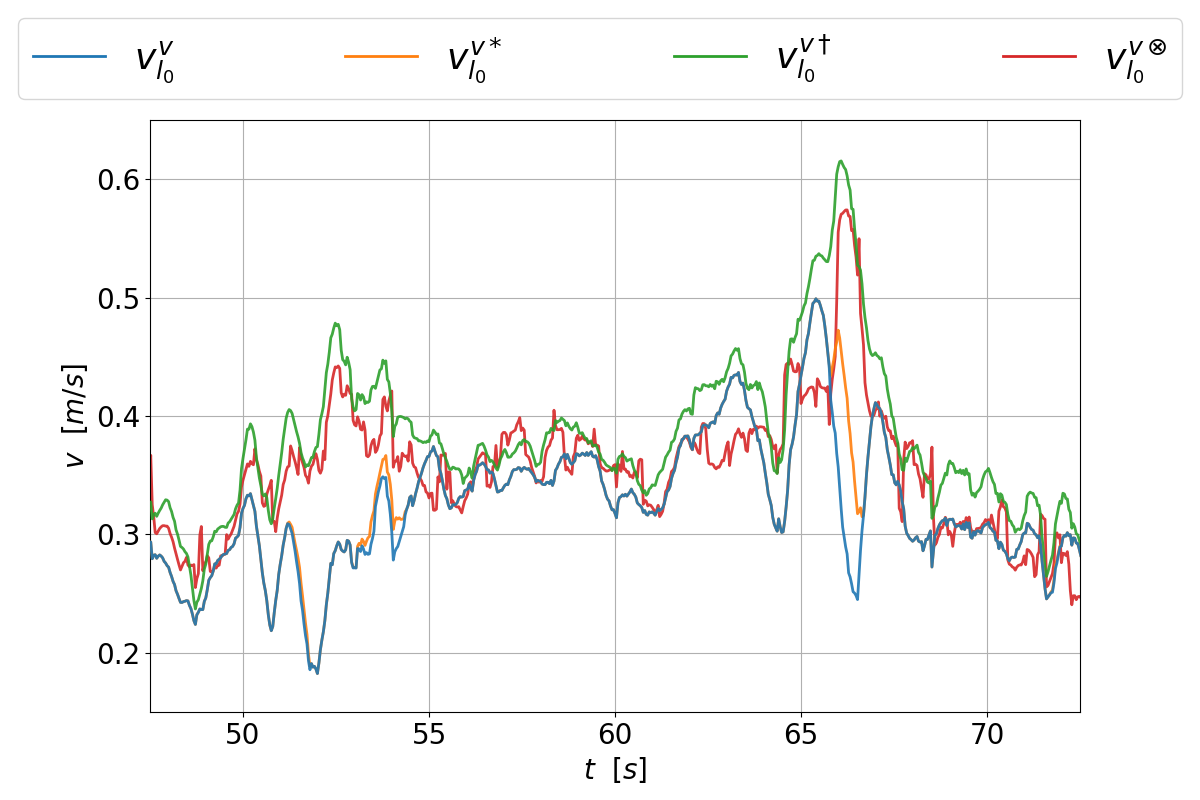}\\
    \includegraphics[width=0.485\textwidth]{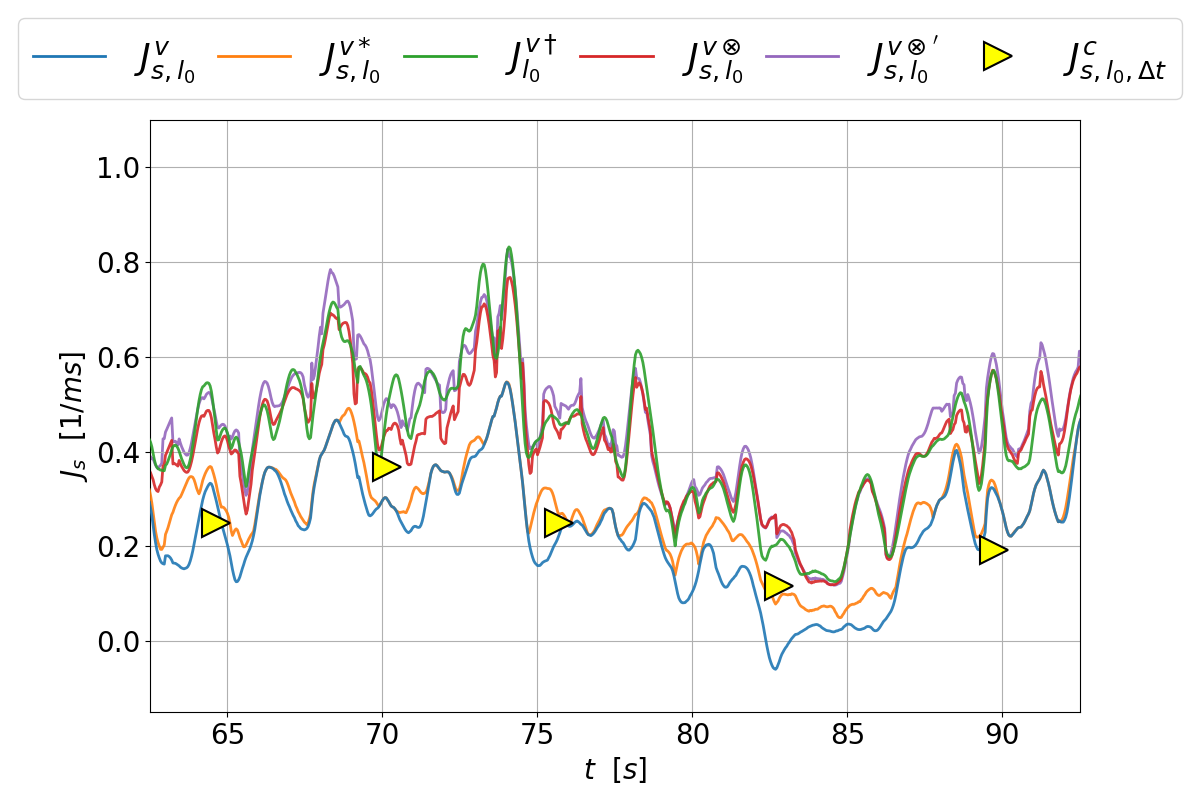}
    \includegraphics[width=0.485\textwidth]{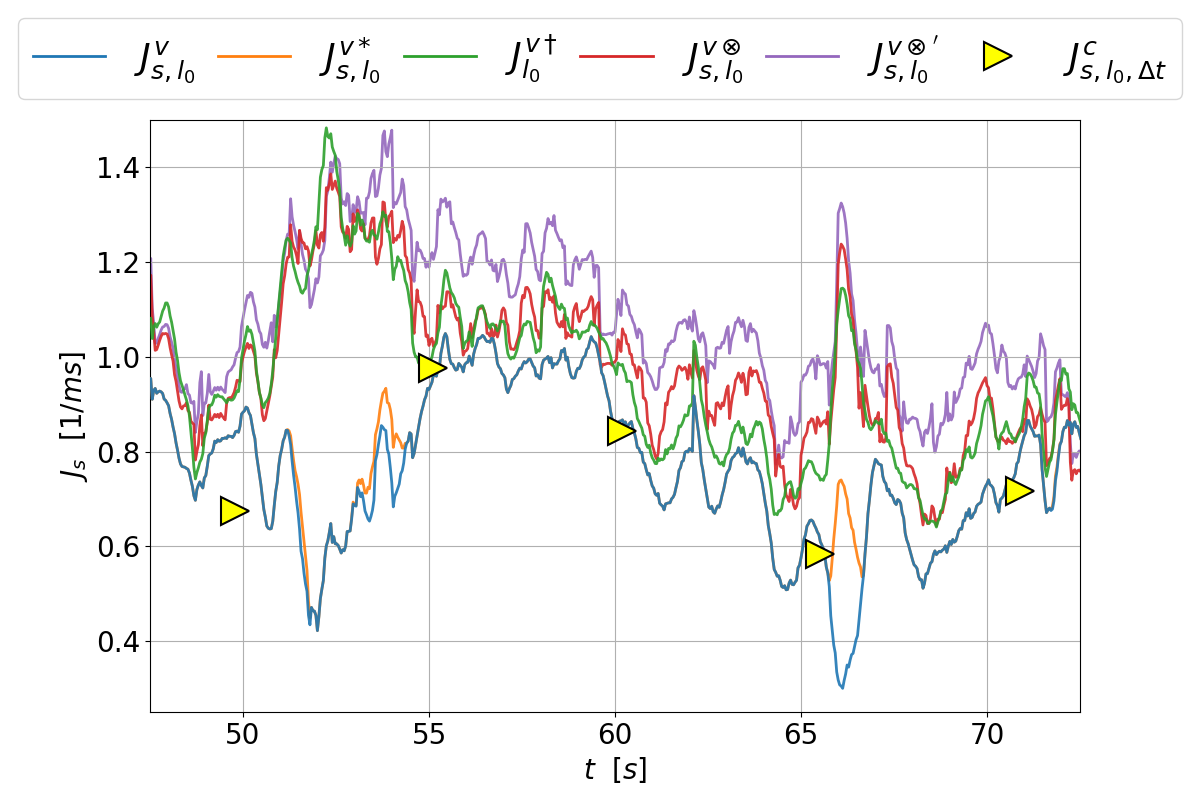}
    \caption{Comparison of time series of density, speed and specific flow applying the different simplifications on corridor experiments. Left column: uni-directional experiment $uni\_corr\_500\_10$; Right column: bi-directional experiment $bi\_corr\_400\_08$.}
    \label{fig:simpli}
\end{figure}

Using the experimental data as an example, we will now examine the extent to which the individual simplifications change the values for density, velocity, and flow. An overview is given in \autoref{tab:J_simplifactions_rms}, summarising the agreement between the flow calculations and the classical flow measurements on the line. This was done by calculating the root mean square (RMS) between the time averages of the calculated data (\autoref{eq:j_mean_delt}) and the classical flow $J_{s,l_0,\Delta t}^{\text{c}}$ (\autoref{eq:variables_flow_macro}) measured at the same time intervals. The RMS was calculated as follows 
\begin{equation}
    RMS = \sqrt{\frac{1}{N} \sum_{n=1}^{N}\frac{\left[J_{s,l_0,\Delta t}^{\text{v}}(\Delta t_n)-J_{s,l_0,\Delta t}^{\text{c}}(\Delta t_n)\right]^2}{\left[J_{s,l_0,\Delta t}^{\text{c}}(\Delta t_n)\right]^2}}
    \label{eq:rms}
\end{equation}
with $N$ being the number of time intervals $\Delta t$ during the experiment. In principle, \autoref{tab:J_simplifactions_rms} illustrate that every simplification leads to an increase in the RMS value and thus to a discrepancy with the classic flow measurement $J_{s,l_0,\Delta t}^{\text{c}}$. The definition $J_{s,l_0}^{\text{v}}$ has the lowest RMS value for all experiments. The high RMS values for the experiment uni\_corr\_10 can be explained by the very low flow near zero (standstill). It should be reconsidered here how to deal with multiple crossings of the measurement line by one person in classic flow measurements, see last sentence in the paragraph below equation \autoref{eq:variables_flow_macro}.

The following comparison of measurement methods is used to analyse the consistency between direct and indirect measurements of the variables, such as $J=\frac{\Delta N}{\Delta t}$ to $J=f(\rho,v)$ in form of time series. The differences in the calculation simplifications are less clear in the experiments with low density and high flux ($uni\_corr\_500\_03$, $bi\_corr\_400\_03$). 
With high density and small flow ($uni\_corr\_500\_10$, $bi\_corr\_400\_08$), however, they are sometimes significant. We therefore focus on these experiments in \autoref{fig:simpli}.\\
In the case of density, the difference caused by using the weighting term are notable. The effect of the weighting is a smoother curve, and the density is smaller (see $\rho_{l_0}^{\text{v}\otimes}$).\\
The order of averaging and multiplication in calculating the specific flow (difference between $J_{s,l_0}^{\text{v}\otimes'}$ and $J_{s,l_0}^{\text{v}\otimes}$) has the greatest influence at high densities. This can be seen in the example of $bi\_corr\_400\_08$, where the mean value $J_{s,l_0}^{\text{v}\otimes'}$ (following \autoref{eq:j_average2}) generally results in higher flow values than the mean value$J_{s,l_0}^{\text{v}\otimes}$ (following \autoref{eq:j_average}). Otherwise, the effect is less pronounced.\\ 
The remaining observations on speed and flow can be summarised together, as the effects of the simplifications on them are similar. 
The difference between $\text{v}\otimes$ and $\text{v}\dagger$ is the  weighting of the contributions of the pedestrian $i$ ($1/N$ or $w_i/w$). The effect is mainly a smoothing (as before with the density). 
Using the normal component of the speed (transition from $\text{v}\dagger$ to $\text{v}\ast$) has the greatest effect. The effect is most evident for experiment $bi\_corr\_400\_08$ between $t=50-55$ s. If $\|\vec{v}\|$ is used ($v_{l_0}^{\text{v}\dagger}$ and $J_{s,l_0}^{\text{v}\dagger}$), the specific flow increases to approximately \SI{1.4}{ms^{-1}}, whereas it drops to almost \SI{0.4}{ms^{-1}} if the normal velocity component $\|\vec{v}\vec{n}\|$ is used ($v_{l_0}^{\text{v}\ast}$ and $J_{s,l_0}^{\text{v}\ast}$). The significance of this effect is also reflected in \autoref{tab:J_simplifactions_rms}.
Finally, versions $\text{v}\ast$ and $\text{v}$ differ only in the use of the magnitude $\|\vec{v}\vec{n}\|$ or the sign of the vertical velocity $\vec{v}\vec{n}$. In most cases, the two calculated velocities ($v_{l_0}^{\text{v}\ast}$ and $v_{l_0}^{\text{v}}$) and specific flows ($J_{s,l_0}^{\text{v}\ast}$ and $J_{s,l_0}^{\text{v}}$) agree very well. However, there are two cases that we would like to draw attention to:\\
1. In the experiment $uni\_corr\_500\_10$ between $t=80-85$ s, a negative velocity $v_{l_0}^v$ and therefore also a negative specific flow $J_{s,l_0}^{\text{v}}$ is measured. This is exactly one of the cases that we wanted to make detectable with the proposed measurement method.\\
2. In experiment $bi\_corr\_400\_08$ at approximately $t=65-70$ s, an increase in velocity is measured for $v_{l_0}^{\text{v}\ast}$, while $v_{l_0}^{\text{v}}$ drops significantly.
After comparing the graphs in \autoref{fig:simpli} and the RMS errors in \autoref{tab:J_simplifactions_rms}, it can be summarised that the calculated values decrease from $\text{v}\otimes'$ to $\text{v}$ and the agreement with the flow measured on the line improves from $\text{v}\otimes'$ to $\text{v}$. The largest step occurs from $\text{v}\dagger$ to $\text{v}\ast$ (as described above).\\

For more information with respect to the full range of the fundamental diagram we refer to   \autoref{fig:fd_uni_mean_nt} and \autoref{fig:fd_bi_mean_nt} where the fundamental diagrams measured with simplified methods are included. 

\section{Summary and Conclusions}

The measurement methods and the method of calculating variables of pedestrian dynamics have not kept pace with the development and the precision of data trajectory recordings.
For reasons of tradition, the flow equation was and is used again and again, even though it has issues as discussed in section 2.2. 
Firstly, the flow equation does not specify how the value of the scalar velocity is calculated from the given velocity vector. Therefore, it is unclear if and how negative contributions to the flow are taken into account. 
Second, density is usually measured as an average over a two-dimensional area. Inhomogeneities are not taken into account and spatial averages (density) are combined with temporal averages (flow). 
Thirdly, multiplication by the length of the measurement line implies that the product of density and velocity along the measurement line is constant. 
Fourthly, when using the flow equation to calculate the flow from measured values of density and velocity, one has to be aware of the Cauchy-Schwartz inequality, which means that the multiplication of the mean values of density and velocity does not necessarily equal the mean value of density times velocity.

The problems listed above show that previous measurements of flow, density and speed measured by trajectories of the head do not necessarily conform to the continuity equation.
While the continuity equation as a conservation law describing changes in density and velocity in time and space, ensuring that no pedestrian is born or turns to dust (conserving the number of pedestrians).
With this problems inaccuracies or uncertainties in the fundamental diagrams can lead to misinterpretation of important crowd conditions (e.g. transport infrastructure capacity or the criticality of conditions in a crowd).

In this paper a framework is introduced that allows the definition of density, speed and flow on the basis of trajectories in accordance with the continuity equation.
In order to relate the continuity equation in it's field representation to the discrete trajectories of the pedestrians, a Voronoi decomposition is applied. A method was introduced to measure density, speed and flow on a line and to deal with different main directions of motion (without losing the sign of the velocity vector).
The newly introduced measurement method have been applied to precise empirical data from laboratory experiments. The results are compared with classical measurements and the effect of various simplifications of the measurement method on the results have been analysed. This shows that there are density ranges in which the influence of the measurement method is small (especially in the free-flow branch). In the high density range, however, the difference is clearly recognisable. This is the density range in which, for example, stop-and-go waves and thus a heterogeneous distribution of density, velocity and flow can occur. 

It should be noted that the method introduced measures the flow perpendicular to a measuring line. In cases where the performance in multidimensional traffic has to bee determined, it must be considered how to place the measuring line in a meaningful way. The number, length and positioning of the measuring lines are completely free and guaranteed to conform to the continuity equation. This method makes it possible to define measuring lines of any length and any complexity at all possible positions in the system. 

In order to further investigate the gain of the newly proposed measurement method, new empirical studies are necessary. These experiments should involve transitions between a moving mode and a stopping mode (and vice versa). The result should be a more accurate dataset of the high density area of the fundamental diagram. This should be applied in uni- as well as bi- and multidirectional pedestrian flows. These data and measurement methods will provide new and precise information on conditions in high-density, like stop and go waves or  pushing crowds where transversal density waves are observable. These findings offer the potential to increase the safety in crowds.

\section*{Acknowledgements}

The authors would like to thank Tobias Schrödter and Christian Hirt for implementing the new measurement method in \href{https://pedpy.readthedocs.io/en/stable/}{PedPy}, \cite{schrodter_2024_11778532}.


\bibliography{bib_fd}

\end{document}